\documentclass[12pt, preprint]{emulateapj}

\usepackage{lscape} 

\shortauthors{Shi et al.}

\begin{document}

\title{BH Accretion in Low-Mass Galaxies Since $z$ $\sim$ 1}

\author{Yong Shi\altaffilmark{1}, George Rieke\altaffilmark{1}, Jennifer Donley\altaffilmark{1},
Michael	 Cooper\altaffilmark{1,7}, Christopher Willmer\altaffilmark{1}, Evan Kirby\altaffilmark{4}}

%Kirpal Nandra\altaffilmark{2},   DONE
%Elise Laird\altaffilmark{2},  DONE
%Marc Davis\altaffilmark{3},  DONE
%Kevin Bundy\altaffilmark{6}, 
%Alison Coil\altaffilmark{1}, 

%Sandy Faber\altaffilmark{4},   DONE
%Puragra Guhathakurta\altaffilmark{4},  DONE
%Evan Kirby\altaffilmark{4},  DONE
%David Koo\altaffilmark{4},   DONE
%Jeffrey Newman\altaffilmark{5},  DONE
%Andrew C. Phillips\altaffilmark{4},  DONE
%David Rosario\altaffilmark{4}}  DONE

\altaffiltext{1}{Steward Observatory, University of Arizona, 933 N Cherry Ave, Tucson, AZ 85721, USA}
%\altaffiltext{2}{Astrophysics Group, Imperial College London, Blackett Laboratory, Prince Consort Road, London SW7 2AZ, UK.}
%\altaffiltext{3}{Astronomy Department, University of California, Berkeley, CA 94720, USA}
\altaffiltext{4}{UCO/Lick Observatory, Department of Astronomy and Astrophysics, University of California, Santa Cruz, CA 95064}
%\altaffiltext{5}{University of Pittsburgh, Pittsburgh, PA 15260, USA}
%\altaffiltext{6}{Department of Astronomy and Astrophysics, University of Toronto, Toronto, ON M5S 3H4, CA}
\altaffiltext{7}{Spitzer Fellow}

\begin{abstract}

We have  selected a  sample of X-ray-emitting active  galactic nuclei
(AGNs)   in   low-mass   host   galaxies   ($\sim$5$\times10^{9}$-
2$\times10^{10}$ M$_{\odot}$)  out to $z$$\sim$1. By  comparing to AGNs
in more massive  hosts, we have found that  the AGN spatial number density
and the fraction of galaxies  hosting AGNs depends strongly on the host
mass, with the AGN host mass function peaking at intermediate mass and
with  the AGN  fraction increasing  with host  mass.  AGNs  in low-mass
hosts  show strong cosmic  evolution in  comoving number  density, the
fraction of such galaxies hosting active nuclei and the comoving X-ray
energy density.   The integrated X-ray luminosity function  is used to
estimate the  amount of the accreted  black hole mass in  these AGNs and
places a strong lower limit of  12\% to the fraction of local low-mass
galaxies hosting black  holes, although a more likely  value is probably
much  higher ($>$50\%)  once the  heavily obscured objects  missed in
current X-ray surveys are accounted for.

\end{abstract}                                                    

\section{Introduction} 

Active  galactic nuclei  (AGNs),  the manifestation  of accretion  onto
massive  black holes  (MBHs),  have   been  recognized  as  a  critical
ingredient in galaxy formation  and evolution. The demography of local
galaxies suggests  that most --  perhaps all -- massive  galaxies host
MBHs  at their centers  and that  MBH masses  are correlated  with the
galaxy  bulge properties  \citep{Kormendy95,  Magorrian98, Gebhardt00,
Ferrarese00, Haring04},  implying the  coevolution of  the galaxy
and MBH.   The good match between  the local BH mass  density and the mass
density of AGN relics further  suggests that all massive galaxies have
experienced an AGN  phase during their  evolution \citep[e.g.][]{Aller02,
Shankar04,  Marconi04}.   Energy  feedback  from  AGNs  to  their  host
galaxies  is invoked to  explain different  aspects of  massive galaxy
evolution.  AGNs may serve as  the heating sources for cooling flows in
clusters \citep[e.g.][]{McNamara07}.  They may suppress star formation
in their host  galaxies and cause them to migrate  from the blue cloud
to  the  red-sequence  in  the color-magnitude  plot  \citep{Croton06,
Nandra07,  Georgakakis08}  and they  may  account for  ``down-sizing''
galaxy  evolution  \citep[e.g.][]{Cowie96}.    As  the  most  abundant
population in the universe,  low-mass (defined as stellar mass $M_{*}$
$<$ 2$\times$10$^{10}$ M$_{\odot}$ through this paper) galaxies act as
the  building  blocks  of  massive  galaxies.   However,  the  current
understanding of  MBHs' role in low-mass galaxy  evolution is limited,
leaving  us with  some basic  questions: Is  the existence  of  BHs in
low-mass galaxies  as common  as it is in massive  galaxies?  Are  there two
types of low-mass galaxy populations  (i.e.  ones with BHs vs. ones
without  BHs)?  How many  low-mass galaxies  experience an  AGN phase?
All of these questions are  related to a more basic  astrophysical problem: what is
the  black hole occupation  function  (BHOF; the  fraction of  galaxies
hosting either active or quiet BHs) in low-mass galaxies?

Searching for  BHs in low-mass systems offers an unique opportunity to
extend the  MBH-$\sigma$ correlation  for massive galaxies,  which not
only tests  the universality of  the relation but  also is
required  to understand the  origin of  the relation.   Although the
MBH-$\sigma$ relationship  has been confirmed for  low-mass systems in
the   local  universe   \citep{Barth05},  some   works   suggest  that
the MBH-$\sigma$ relationship  may be  replaced at low  mass by  a similar
relation    involving   compact   stellar    nuclei   \citep{Wehner06,
Ferrarese06, Rossa06}.

A  better understanding  of primordial  MBH seed  growth in  the early
universe may also benefit from  the study of BHs in low-mass galaxies,
as  such work  could reveal  aspects of  the accretion  mode in  a low
gravitational  potential and  low metallicity  environment and  of the
relative  importance  of mass  growth  through  accretion and  merging
processes. For example, in  low-mass galaxies, the impulsive kick from
anisotropic gravitational emission during  BH-BH mergers is thought to
be  strong enough  to eject  central BHs  \citep{Favata04, Merritt04}.
Understanding AGN activity in  low-mass galaxies can further constrain
MBH  seed formation  theories.  In  a cold  dark matter  universe, the
primordial  MBH  seeds  form  as  remnants  of  Population  III  stars
\citep[e.g.][]{Madau01} or through the direct collapse of pre-galactic gas
disks  \citep{Lodato06}.   The  efficiencies  in  different  formation
scenarios predict a wide range  in the local BHOF in low-mass galaxies
which can vary from zero to unity, while the prediction of the BHOF in
massive galaxies  is invariably  unity as constrained  by observations
\citep{Volonteri08}.

The current study of BHs in low mass galaxies is mostly limited to low
redshift ($z$  $<$ 0.3).  Despite  a variety of investigations,  it is
still  unclear  how many  local  low-mass  galaxies  harbor BHs.   The
dynamical searches  for MBHs have confirmed  the existence of  a BH in
M32 \citep{Verolme02}, but not in M33 \citep{Merritt01, Gebhardt01} or
NGC205 \citep{Valluri05}.  Searching  for actively accreting BHs (i.e.
AGNs)  has provided  a stronger  lowerlimit and  more complete  view of
local BHOFs as a function  of the system mass through optical emission
lines  \citep[e.g.][]{Filippenko89,  Maiolino95,  Ho95,  Kauffmann03a,
Greene04,     Greene07c,     Decarli07, Dong07, Shields08}, mid-infrared line 
selections \citep{Satyapal07, Satyapal08}     and     X-ray     emission
\citep[e.g.][]{Gallo07, Ghosh08}.  It has been shown that the fraction
of  galaxies hosting active  BHs depends  on both  BH mass  and galaxy
stellar mass, peaking at an intermediate mass range and falling toward
higher and lower mass \citep{Kauffmann03a, Heckman04, Greene07b}.

As the local  BH masses are most likely  accumulated at high redshift,
the  search for  AGN activity  in low-mass  galaxies at  high redshift
should  provide independent  and  possibly better  constraints on  the
local BHOF. Such study is also significant for understanding the evolution
of  nuclear activity  in low-mass  galaxies and  the role  of  AGNs in
low-mass galaxy evolution.  With the  advent of deep {\it Chandra} and
{\it XMM-Newton}  X-ray surveys, we  have searched for  X-ray emitting
AGNs in low-mass host galaxies out to $z$ $\sim$ 1.  In this paper, we
first describe  the identification of  AGNs in low-mass  galaxies (see
\S~\ref{DATA} and  \S~\ref{DAGN_DEF}).  We then  present the 1/$V_{\rm
max}$ method  to correct for  incompleteness in \S~\ref{Method_NumDen}
and study  their spatial number density and  X-ray luminosity function
in  \S~\ref{RESULT}.  In \S~\ref{Discussion},  we discuss  the local
BHOF in  low-mass galaxies constrained  by our study  of high redshift
AGNs  in   such  galaxies.   Our  conclusions   are  presented  in
\S~\ref{Conclusions}.  Throughout  this paper, ``low-mass''  refers to
normal galaxies  or AGN  host galaxies with  stellar mass  $M_{*}$ $<$
2$\times$10$^{10}$  M$_{\odot}$ and  ``massive'' indicates  those with
stellar mass  $M_{*}$ $>$ 2$\times$10$^{10}$ M$_{\odot}$.   We adopt a
cosmology with $H_{0}$=70 km s$^{-1}$ Mpc$^{-1}$, $\Omega_{\rm m}$=0.3
and  $\Omega_{\Lambda}$=0.7.  All  magnitudes  are defined  in the  AB
system.

%Finally,  we discuss their  X-ray continuum  properties to  give the  first glimpse  of the
%accretion in these low mass systems. 

\section{Data: X-ray Survey Fields}\label{DATA}

We have searched for AGNs in  low-mass host galaxies in five {\it Chandra}
and  {\it XMM-Newton}  fields, including  the  All-Wavelength Extended
Groth  Strip  International Survey  (AEGIS),  the  {\it Chandra}  Deep
Field-North Survey (CDF-N), the  {\it Chandra} Deep Field-South Survey
(CDF-S), the  {\it Chandra} Large-Area Synoptic  X-Ray Survey (CLASXS)
and  the  {\it  XMM-Newton}  Large Scale  Structure  Survey  (XMMLSS).
Table~\ref{xrayfields}  lists  the properties  of  these five  fields,
including  the  area, the  limiting  hard  X-ray  flux, the  available
optical/near-IR photometry,  the definition of  secure (multiple-line)
spectroscopic redshifts and the associated references.

All the X-ray fields  are fully covered by optical/near-IR photometry.
For  the CDF-N,  CDF-S and  CLASXS fields,  spectroscopic observations
have been  obtained for  X-ray sources, while  the redshifts  of X-ray
objects in  the remaining  two fields are  obtained by  matching X-ray
catalogs to galaxy redshift survey catalogs.  Table~\ref{xrayfield_LF}
summarizes  the total  number of  X-ray sources  and  of spectroscopic
targets.   The search  radii  for optical  counterparts  to the  X-ray
sources  are  2.0$''$ and  2.5$''$  for  the  {\it Chandra}  and  {\it
XMM-Newton}  fields,  respectively.    The  optical  counterparts  are
identified  using  $R$-band  catalogs,  which  generally  provide  the
deepest  observations.  If  multiple optical  objects within  a search
aperture  are present,  the  closest  one is  defined  as the  optical
counterpart.   The fraction  of  X-ray objects  with multiple  optical
sources within a  single aperture is only $\sim$  10\%. Therefore, the
assumption  of adopting  the closest  one as  the  optical counterpart
should  not affect  our conclusions.   We  have limited  our study  to
objects  with secure  spectroscopic redshifts  (see \S~\ref{DAGN_DEF})
and thus the majority of the optical counterparts are brighter than 24
in the $R$-band.  At $R<$24, the surface density of galaxies and stars
is  about  16.6 arcmin$^{-2}$  \citep{Capak04}.   The probability  for
chance superposition between X-ray and optical sources is 6\% and 10\%
for  the  {\it Chandra}  and  {\it  XMM-Newton} fields,  respectively.
Given a total of 32 low-mass AGN hosts in the {\it Chandra} fields and
zero in the {\it XMM-Newton} field, we have estimated that only two objects
are expected  to have spurious optical counterparts.   Our sample only
includes the X-ray sources with detected hard X-ray fluxes, defined in
the energy  range of 2-8  keV.  The published  2-10 keV fluxes  in the
AEGIS and  XMMLSS fields  have been corrected  to 2-8 keV on the assumption of 
a power-law  photon  index  of  1.0,  which is  the  average  of
hard-X-ray  selected   objects    based   on   the   hardness   ratio
\citep{Nandra05}.

All  optical type  1  AGNs  are excluded  as  their nuclear  radiation
contaminates the  host optical/near-IR light severely. Due  to lack of
access to the observed spectra,  the definition of optical type 1 AGNs
is not completely universal over  all the fields. In the CLASXS field,
type 1  objects are defined by  having an emission line FWHM  $>$ 1000 km
s$^{-1}$, while FWHM $>$ 2000 km  s$^{-1}$ is adopted for the CDF-N and
CDF-S fields.  For  the AEGIS and XMMLSS field,  we have downloaded the
spectra and classified type 1 objects using FWHM $>$ 1000 km s$^{-1}$.
In the AEGIS field, some fraction  of type 1 objects is still included
in  the sample  as the  DEEP2 spectral  coverage misses  the permitted
lines  (H$\alpha$,  H$\beta$ and  MgII2800$\AA$)  in certain  redshift
windows  ($\sim$0.2-0.3  and $\sim$0.6-1.2).  Given  a  broad line  AGN
fraction of  16\% within broad-line-detectable  redshift ranges (0-0.2
and  0.3-0.6)  and  a  total  of  eight  AGNs  in  low-mass  hosts  within
broad-line-undetectable redshift  ranges in the AEGIS  field, only one
object in our final sample may be misclassified as an optical type 2 AGN.

%As low  X-ray luminosity
%AGN  are dominated  by  type 2  objects \citep[e.g.][]{Barger05},  the
%differences  among the  above  classifications should  not affect  our
%results significantly.

\section{Selection of  AGNs in Low-Mass Host Galaxies}\label{DAGN_DEF}

Objects are  identified as active low-mass galaxies if they  satisfy the
following two criteria:  1.)  stellar mass $M_{*}<$ 2$\times$10$^{10}$
M$_{\odot}$, our  definition of {\it  low-mass} galaxies and  2.)  
hard  X-ray luminosity  $L_{\rm  2-8keV}$ $>$  10$^{42}$ erg  s$^{-1}$, indicating an {\it active} nucleus.

We have measured stellar masses  by comparing the observed SEDs (i.e.,
the photometric  data included in  Table~\ref{xrayfields}) to 102168
stellar  synthesis models  produced by  \citet{Bruzual03}'s  code.  As
listed in  Table~\ref{BC_model}, the stellar models span  a wide range
of parameter space,  including metallicity, extinction, characteristic
timescale of  exponential star formation history,  fraction of ejected
gas  being  recycled and  galaxy  age.   To  account for  the  possible
existence of low metallicity and  young galaxies, we have included all
six  available  metallicities and  galaxy  ages  starting at  10$^{6}$
yrs. The fit algorithm is  similar to that of \citet{Bundy06} who have
used a  Bayesian technique  as described in  \citet{Kauffmann03c}.  In
summary,  for  an individual  object,  the  best-fit  stellar mass  is
obtained for  each model that corresponds  to an age  younger than the
cosmic age at  the redshift of the object.   The associated $\chi^{2}$
gives the probability  (exp(-$\chi^{2}$/2)) that this model represents
the observed  SED.  The final probability distribution  of the stellar
mass is  obtained by summing  all probabilities within a  certain mass
bin. We  adopted a bin width  of 0.01 in log$M_{*}$,  which on average
contains 300 models.  The median value of the stellar mass probability
distribution is adopted as the final mass of a galaxy. Compared to the
minimum  $\chi^{2}$ derived stellar  mass, the  mass obtained  by this
technique suffers much less from  model degeneracies.  The 68\%
uncertainty range of the derived  stellar mass is defined by excluding
the 16\% tail at each end of the probability distribution.  Note
that this uncertainty mainly reflects the photometric errors.  For the
objects  in  the CDF-N  and  CLASXS  fields,  there are  no  published
photometric  errors.  We  have adopted  universally an  uncertainty of
0.07 magnitude, which is roughly the sky noise for a $m_{R}$=24 object
in these two fields.  A small  fraction of objects in AEGIS, CDF-S and
XMMLSS with  very small photometric  errors show large  minimum reduced
$\chi^{2}$, indicating our stellar models  are not able to produce the
observed  SEDs  accurately.  To  estimate  the  uncertainty for  these
objects, we increased their photometric errors to 0.02 magnitude, which
gives a reasonable minimum $\chi^{2}$ ($<$ 10) and larger uncertainty.

\begin{figure}
\epsscale{1.0}
\plotone{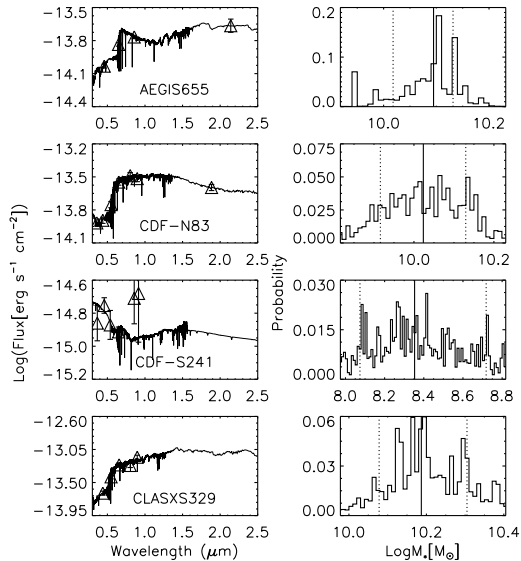}
\caption{ \label{SED} Examples of the best-fit SEDs and the probability distribution
of the stellar mass of AGNs in low-mass galaxies in each field, where the solid line is the 
median stellar mass
and two dotted lines indicate the 16 \% and 84 \% probability tails, respectively.}
\end{figure}

We do  not include  other systematic errors  for our  measured stellar
masses, such as the accuracy  of the \citet{Bruzual03} code itself and
possible  alternative  choices for  the  initial  mass function.   Our
sample  only  includes    narrow-line  AGNs  and  thus  AGN  light
contamination to the host emission should be small. For obscured AGNs,
the scattered nuclear light may be strong \citep{Zakamska06}. However,
such scattered emission may affect  the host light importantly only in
the UV band.  To avoid the dust emission from AGN  dusty tori, we have
not   used  photometry  in the mid-IR.
Fig.~\ref{SED} shows examples of  the observed SED superposed with the
minimum-$\chi^{2}$ stellar  model and the  probability distribution of
the stellar mass for each field.

A mass cut of $<$  2$\times$10$^{10}$ M$_{\odot}$ is adopted to define
the sample  of low-mass host  galaxies.  Selecting the sample  at this
mass threshold is  of interest because the properties  of galaxies and
BHs may transition around this mass.  Studies of low-redshift galaxies
have shown  that galaxy properties  (star formation history,  size and
internal  structure)  show  significant  differences at  the  dividing
stellar  mass of 3$\times$10$^{10}$  M$_{\odot}$ \citep{Kauffmann03b}.
In addition, galaxies with stellar mass $<{\sim}$10$^{10}$ M$_{\odot}$
often harbor compact stellar nuclei at their centers, which follow the
MBH  mass-bulge relationships,  but  which are  rare  in more  massive
galaxies \citep{Carollo98, Laine03, Wehner06,  Ferrarese06}.  These stellar  nuclei may be
replacements  for MBHs  in  low-mass systems  where the  gravitational
potential is  not sufficiently  deep to form  a BH.   Furthermore, the
fraction of galaxies hosting active  BHs depends on the galaxy stellar
mass \citep{Kauffmann03a, Heckman04,  Greene07b}. In the most complete
local AGN sample, the AGN  fraction in galaxies fainter than $M_{B}$ =
-20 ($\sim$10$^{10}$M$_{\odot}$) is on average half of the fraction in
brighter ones \citep{Ho97}.

\begin{figure}
\epsscale{1.0}
\plotone{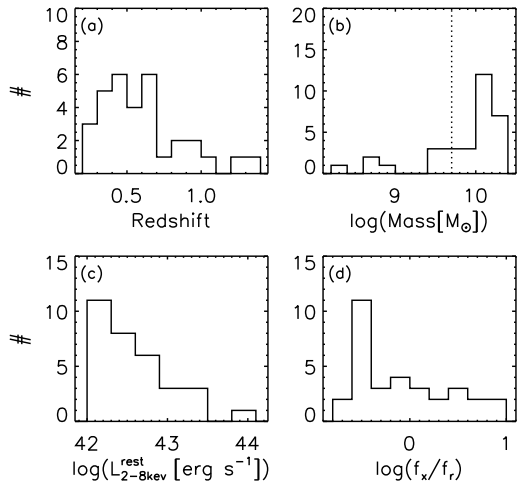}
\caption{ \label{dist_properties}  Distributions of properties  of AGNs
in   low-mass   galaxies:  (a)--   redshift   distribution;  (b)   --
distribution  of host  stellar masses where the dotted line indicates
the completeness cut of 10$^{9.7}$   M$_{\odot}$;  (c) --  distributions  of the
rest-frame 2-8 keV X-ray luminosity;  (d) -- distribution of X-ray to
$R$-band flux ratio. To be complete, our number density and X-ray luminosity
function for AGNs in low-mass hosts are measured for those at 0.1 $<$ z
$<$   0.7,  10$^{9.7}$   M$_{\odot}$  $<$   $M_{*}$   $<$  10$^{10.3}$
M$_{\odot}$ and $L_{\rm 2-8kev}^{\rm rest}$ $>$ 10$^{42}$ erg s$^{-1}$.}
\end{figure}

The  $L_{\rm  2-8keV}$  $>$  10$^{42}$  erg  s$^{-1}$  AGN
selection criterion is mainly based on the energy budget argument that
star-forming galaxies rarely produce such high hard X-ray luminosities
\citep[e.g.][]{Zezas98}.   The  resulting sample  of  AGNs in  low-mass
hosts contains 32 objects as listed in Table~\ref{target}. Each object
is labeled  by the field name  followed by the sequence  number in the
X-ray   catalog  of   each  field   (see   Table~\ref{xrayfields}  for
references).    Fig.~\ref{dist_properties}   shows  distributions   of
low-mass AGN  properties, including  redshift, host stellar  mass, 2-8
keV rest-frame luminosity and X-ray to $R$-band flux ratio.

\citet{Best05} have measured average ratios of BH masses to the galaxy
stellar mass as  a function of stellar masses for  the SDSS galaxy and
AGN samples (see  their Fig.1), where the BH  mass is measured through
the stellar  veloclity dispersion  and the MBH-$\sigma$  relation.  By
assuming that our X-ray selected AGN  sample has the same BH-to-galaxy mass
ratio  as  the SDSS  AGN,  we can  estimate  the  Eddington ratio  for
galaxies with $M_{*}$=  2$\times$10$^{10}$ M$_{\odot}$ and $L_{2-8 \rm
keV}$=10$^{42}$  erg  s$^{-1}$.   By assuming  the  $L_{\rm
bol}/L_{2-10    \rm     keV}=17(L_{2-10    \rm    keV}/10^{43}$    erg
s$^{-1})^{0.43}$  \citep{Shankar04} and  $L_{\rm 2-8  keV}/L_{2-10 \rm
keV}$=0.86  (assuming a power  law photon  index of  1.0), we  have an
Eddington  ratio  of   0.007.   As  shown  in  \S~\ref{Method_NumDen},
although we  can detect  galaxies with stellar  masses down  to around
10$^{8}$ $M_{\odot}$, our  sample is complete down to  stellar mass of
10$^{9.7}$ M$_{\odot}$ out to redshift  of 0.7. For this stellar mass,
the  black  hole  mass  is on  average  3$\times$10$^{6}$  M$_{\odot}$
\citep{Best05}  and  the corresponding  Eddington  ratio  is 0.02  for
$L_{2-8 \rm  keV}$=10$^{42}$ erg s$^{-1}$.   As a comparison
to those  local low-mass AGNs,  our sample may  be not deep  enough to
include the  local classical low-mass Seyfert galaxies  similar to NGC
4395 or POX  52, but should include galaxies  similar to some low-mass
SDSS AGNs found by \citet{Greene04, Greene07c}.

To better understand the role  of AGNs in low-mass galaxy evolution, we
will compare  the sample of  low-mass host AGNs  to a sample of  AGNs in
massive  host galaxies. The  comparison sample  is defined  as stellar
mass  $M_{*}$ $>$  2$\times$10$^{10}$ M$_{\odot}$  and  $L_{\rm
2-8keV}$  $>$ 10$^{42}$  erg s$^{-1}$.

%\subsection{Individual Properties}

\section{Incompleteness Corrections and Weights}\label{Method_NumDen}

\subsection{The 1/$V_{\rm max}$ method}\label{VmaxMethod}

The spatial number density of AGNs as a function of AGN host mass and the
X-ray  luminosity function of  AGNs in  low-mass hosts  was calculated
using the  1/$V_{\rm max}$ method \citep{Schmidt68}.   This calculation used
those objects  with $R$  $<$ 24,  $L_{\rm 2-8keV}^{\rm
rest}$  $>$   10$^{42}$  erg   s$^{-1}$ and   $M_{*}$  $>$
$5\times10^{9}$ M$_{\odot}$  and in two  redshift intervals of 0.1  $<$ $z$
$<$ 0.4 and  0.4 $<$ $z$ $<$ 0.7.  $L_{\rm  2-8keV}^{\rm rest}$ is the
rest-frame  2-8  keV flux  assuming  a photon  index  of  1.0 but  not
correcting for  extinction.   The maximum  volume over which  an object
is included in the sample is given by
\begin{equation}   
V_{\rm   max}    =   \int_{z_{\rm   low}}^{z_{\rm high}} \Omega \frac{dV}{dz} dz,
\end{equation} 
 where  [$z_{\rm  low}$,  $z_{\rm  high}$]  is the  redshift  range  of
interest  and  $\Omega$  is  the  solid angle  covered  by 
the X-ray survey at the  flux level of the object. While $z_{\rm  low}$ is always  
fixed to the  low end of  a redshift interval, the maximum redshift, $z_{\rm high}$ is defined as:

\begin{equation} 
z_{\rm high}={\rm min}(z_{\rm bin}^{\rm high}, z^{\rm limit}_{\rm xray}, z^{\rm limit}_{R}),
\end{equation} 
 where $z_{\rm bin}^{\rm  high}$ is the high end  of a redshift
interval,  $z^{\rm limit}_{\rm xray}$ is the limiting redshift at which the
observed X-ray flux reaches the limiting flux in a given field where the
K-correction  is determined  by  assuming a  power  law spectrum  with
photon  index of 1.0 and  $z^{\rm limit}_{R}$  is the  limiting redshift
where the observed $R$-band magnitude reaches $m^{\rm limit}_{\rm
R}$=24.   To  determine the  K-correction  in  the  $R$-band, we  have
redshifted the  minimum reduced $\chi^{2}$ spectrum  model produced in
our  stellar mass calculations  to measure  the $R$-band  magnitude at
different distances.

\begin{figure}
\epsscale{1.0}
\plotone{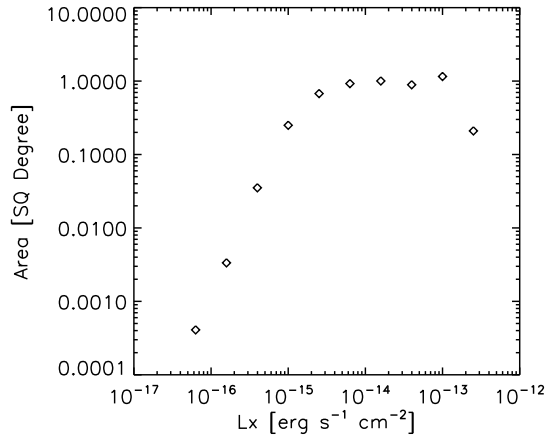}
\caption{ \label{Area_Fx}  The effective solid  angle as a  function of
the  hard X-ray flux  for objects  in AEGIS,  CDF-N, CDF-S  and CLASXS
fields. }
\end{figure}

As the sensitivity of the  X-ray telescope shows energy and positional
dependence  \citep[for details,  see][]{Yang04}, to  obtain  a simple
estimate  of the  solid angle  at  a given  hard X-ray  flux, we  have
followed \citet{Barger05}  by comparing  the observed number  of X-ray
objects at different  hard X-ray fluxes with the  average X-ray number
counts  from  \citet{Cowie02}  and  \citet{Yang04}.   The  sample  for
comparing number counts is  composed of all spectroscopically observed
(not only those  with secure redshifts) X-ray objects  in the field of
CDF-N,  CDF-S and CLASXS  and all  X-ray sources  in the  AEGIS field.
Table~\ref{xrayfield_LF}  lists the  number of  X-ray objects  and the
number of spectroscopic targets  in all the fields.  The spectroscopic
targets  are  randomly  selected  for  the CDF-S  and  CLASXS  fields.
Although the  target selection in the  CDF-N field has  a little bias,
the majority  of sources (439  out of 503)  have been included  in the
spectroscopic sample, making the effect of this bias on the solid
angle measurement negligible \citep{Barger05}.  Note that we have used
the \citet{Alexander03}  X-ray catalog  in the CDF-S  due to  its high
X-ray positional  accuracy so that our catalog  contains $\sim$30 less
objects  than  in \citet{Barger05}.   The  AEGIS  field contains  1318
objects and the redshift is  obtained by matching the X-ray catalog to
the  DEEP2/AEGIS catalog  whose target  selection depends  on apparent
magnitude and color. Therefore, for the objects in the AEGIS field, we
have  applied weights  as shown  below.  Fig.~\ref{Area_Fx}  shows the
effective solid  angle as a function  of the hard X-ray  flux.  At the
flux  of 10$^{-14}$  erg s$^{-1}$  cm$^{-2}$, it  is dominated  by the
CLASXS (0.4 square degree) and AEGIS fields (0.67 square degree).

We  have followed  the method  in \citet{Willmer06}  to  determine the
galaxy weights,  $\omega$, for the  X-ray objects in the  AEGIS field.
These  weights are  used  to  account for  the  under-sampling of  the
photometric catalog (i.e.  the  fraction of objects with spectroscopic
observations)  and  the redshift  success  rate  of the  spectroscopic
catalog  (i.e.   the fraction  of  spectroscopic  targets with  secure
redshift) of the DEEP2 observations.  Briefly, we  define a X-ray-optical
photometric  catalog as  all  AEGIS X-ray  objects  that have  optical
counterparts  in  \citet{Coil04}.  Here  we  assume  that AEGIS  X-ray
objects without  optical counterparts (about  25\% of the  whole X-ray
sample) are not  of interest for our study of  low-mass systems at $z$
$<$ 0.7, either at higher redshift or with stellar mass lower than our
completeness limit  (10$^{9.7}$ M$_{\odot}$). To  demonstrate this, we
found that 15\% of  the X-ray objects without optical counterparts 
have $L_{\rm x}$ $>$ 10$^{42}$ erg s$^{-1}$ at z=0.7, i.e. satisfy the AGN
definition. However, none of  these objects have absolute $R$-band
magnitude  for $m_{R}$  =  24.75 at  z=0.7  brighter than  any of  the
observed low-mass  AGN hosts  with 9.7 $<$  Log($M_{*}/M_{\odot}$) $<$
10.3, where $m_{R}$ = 24.75  is the completeness cut of the DEEP2 photometry
catalog  \citep{Coil04}.   For each  AEGIS  X-ray  object with a  secure
spectroscopic redshift and probability of being a galaxy $P_{\rm gal}$
$>$    0.2,    a    data    cube    in    the    magnitude-color-color
($R$-($B$-$R$)-($R$-$I$))  space is  defined. Then  a  probability for
being within the permitted redshift limit [0.1, 1.4] is calculated for
each object in the X-ray  optical photometric catalog within this data
cube. The sum of all  these probabilities within the data cube divided
by the number of successful redshifts gives the weight $\omega$, which
is further corrected  for the probability of being  placed in the slit
mask.

\begin{figure}
\epsscale{1.0}
\plotone{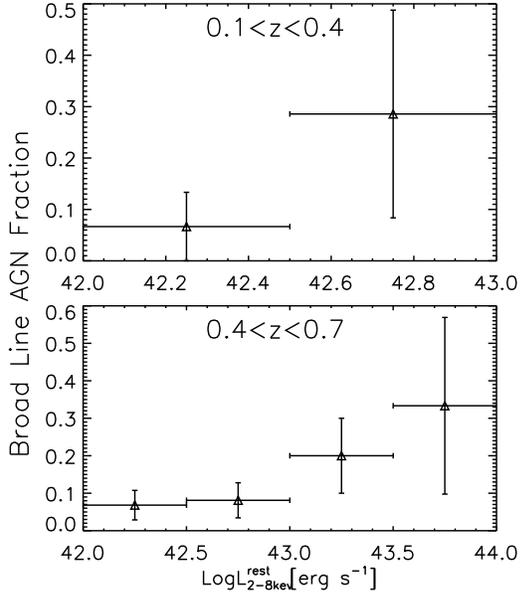}
\caption{ \label{BLAGNFRAC} The broad line AGN fraction in two redshift
intervals of 0 $<$ $z$ $<$ 0.4  and 0.4 $<$ $z$ $<$ 0.7.}
\end{figure}

As  discussed in  \S~\ref{DATA}, type  1  AGNs are  excluded from  the
sample and  thus a  weight is applied  to correct for  their omission.
The fraction of  type 1 AGNs as a function of  the X-ray luminosity is
constructed    in    both    redshift    intervals   as    shown    in
Fig.~\ref{BLAGNFRAC}, which  is almost the same as  the result obtained
by \citet{Barger05}.   The weight is calculated for  each object based
on  the  X-ray  luminosity.  As  shown  in  Fig.~\ref{BLAGNFRAC},  the
corrections are small,  as our AGN sample does  not extend toward high
X-ray luminosity  where the broad  line fraction is  large. Therefore,
our results  throughout the paper  do not change significantly  if the
broad-line AGN fraction does not apply.

Finally, the mass  function and X-ray luminosity function are
determined as:

\begin{equation}
\Phi(X)dX = \sum \omega/V_{\rm max} dX,
\end{equation}
 where    $X$   is    the   stellar    mass   or    X-ray   luminosity,
respectively, $\omega$ is  the galaxy weight, and $V_{\rm  max}$ is the
maximum volume for its detection. 

%The resulting mass functions for the two redshift intervals
%are illustrated  in Fig.~\ref{LF_MHost}. The   luminosity
%functions of AGN in low-mass hosts  are shown   in Fig.~\ref{LF_xray}.

\subsection{Incompleteness}

\begin{figure}
\epsscale{1.0}
\plotone{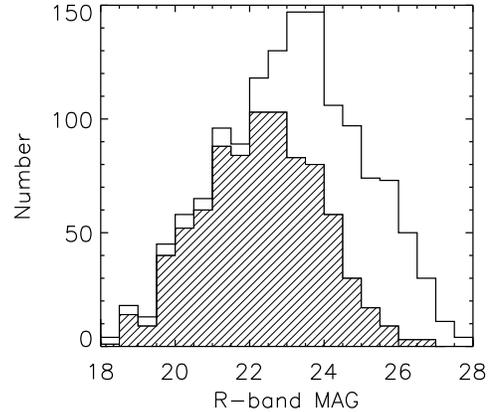}
\caption{ \label{spec_complete} The number distribution of all spectroscopic targets (open)
and objects with secure spectroscopic redshift (hatched).}
\end{figure}

Our AGN  sample where the  1/$V_{\rm max}$ method applies  are objects
with   secure    spectroscopic   redshifts,   $R<$24,    $M_{*}$   $>$
$5\times10^{9}$  M$_{\odot}$,  and  $L_{\rm  2-8keV}^{\rm  rest}$  $>$
10$^{42}$  erg   s$^{-1}$.   The  criteria   of  secure  spectroscopic
redshifts   does  not   introduce   any  significant   incompleteness.
Fig.~\ref{spec_complete}  shows the  distribution of  R-band magnitude
for the spectroscopically observed  X-ray objects (open histogram) and
the  objects with  secure spectroscopic  redshifts  (filled histogram).
The  redshift success  rate  is 76\%  and  68\% for  spectroscopically
observed  objects at  $R$ $<$24  and 22  $<$ $R$  $<$24, respectively.
Most of  the spectroscopically failed  objects have 22 $<$  $R$ $<$24.
\citet{Barger05}   have   shown   that   most  objects   with   failed
spectroscopic  redshifts   have  photometric  redshifts   larger  than
$\sim$1.  A similar  result is  found in  DEEP2  \citep{Willmer06}.  A
rough  estimate shows  that all the spectroscopically  failed  objects contain
roughly  only  two low-mass  AGN  hosts at  $z$  $<$  0.7, given  that
$\sim$25\%  of the spectroscopically  failed objects  with 22  $<$ $R$
$<$24 have a photometric redshift smaller than 0.7 and that 5\% of the
objects with secure redshift with 22 $<$ $R$ $<$24 are of low mass.

The limiting  $R$-band magnitude of 24  is deep enough  to sample most
low-mass galaxies with $M_{*}$  $>$ $5\times10^{9}$ M$_{\odot}$ out to
a redshift of 0.7, the  upperlimit where the spatial number density is
measured.   This  is  because  80\%  of  AGNs  with  host  $M_{*}$  $>$
$5\times10^{9}$ are detectable beyond a redshift of 0.7.  The omission
of a small  fraction of low-mass hosts should  not affect the comoving
number density as they have  been accounted for in the 1/$V_{\rm max}$
method.  To further  demonstrate  that the $R<$ 24 limit  does  not introduce  any
significant incompleteness even in the high redshift interval (0.4 $<$
$z$ $<$ 0.7),  we carried out a Monte-Carlo  simulation to demonstrate
that if the comoving number density is constant within 0.4 $<$ $z$ $<$
0.7, the  1/$V_{\rm max}$ method can re-produce  the intrinsic spatial
density, regardless  of the loss of  a small fraction  of faint R-band
sources.   Briefly, we  randomly populated  the volume  with AGNs  to a
total number similar to that observed  for 0.4 $<$ $z$ $<$ 0.7.  Their
SEDs are assigned randomly  to be one of those at 0.1  $<$ $z$ $<$ 0.4
where the limiting $R$-band magnitude of 24 should detect all low-mass
galaxies  with  $M_{*}$ $>$  5$\times$10$^{9}$  M$_{\odot}$.  This  is
because the  \citet{Bruzual03} oldest simple  stellar populations with
solar metallicity  and $M_{*}$ $=$  5$\times$10$^{9}$ M$_{\odot}$ have
$m_{R}$=24 at  z=0.4. Younger   or lower metallicity  galaxies will
emit  higher  luminosities.   We  then applied  the  measured  $z^{\rm
limit}_{R}$   to the simulated  galaxies   and  measured   the  1/$V_{\rm
max}$-based spatial number density.   A thousand simulations show that
there  is no  systematic difference  in  our estimate  of the  spatial
density. For the incompleteness  due to the X-ray detection threshold,
since all  AGNs (defined as  $L_{\rm 2-8keV}^{\rm rest}$  $>$ 10$^{42}$
erg s$^{-1}$) in low-mass  hosts with $M_{*}$ $>$ $5\times10^{9}$ have
$z^{\rm limit}_{\rm Xray}$ $>$  0.7, the 1/$V_{\rm max}$ method should
recover the real spatial number density.

\section{RESULTS}\label{RESULT}

\subsection{Comoving Number Density and AGN Fraction}\label{COMOVING_ND}

\begin{figure}
\epsscale{1.0}
\plotone{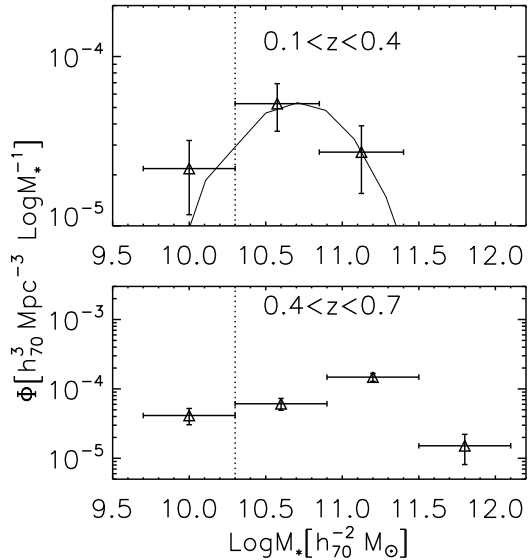}
\vspace{0.3cm}
\caption{ \label{LF_MHost} The stellar mass function of AGN hosts with
$L_{2-8keV}^{\rm rest}$ $>$ 10$^{42}$ erg s$^{-1}$ in the two redshift
intervals of  0.1 $<$  $z$ $<$ 0.4  and 0.4  $<$ $z$ $<$  0.7.  Dotted
lines show the dividing mass between low-mass and massive hosts.
The  solid  line in  the  upper panel shows the SDSS AGN  result at $z$ $<$ 0.3 obtained
by   multiplying   the  fraction   of   galaxies   hosting  AGNs   from
\citet{Kauffmann03a} with the galaxy mass  function at 0.2 $<$ $z$ $<$
0.4 from  \citet{Perez-Gonzalez08}. The normalization of the AGN fraction of \citet{Kauffmann03a} has been
decreased by  a factor of 8  to match our  normalization (a discussion
about  the difference  between the  SDSS   and  our low-redshift
samples  can be found  in \S~\ref{red_DAGN}). }
\end{figure}

Fig.~\ref{LF_MHost}  shows  the spatial  number  density  of AGNs  with
$L_{\rm 2-8keV}^{\rm  rest}$ $>$ 10$^{42}$ erg s$^{-1}$  as a function
of AGN host mass in the two  redshift intervals of 0.1 $<$ $z$ $<$ 0.4
and 0.4 $<$ $z$ $<$ 0.7.   Dotted lines show the dividing mass between
low-mass and massive hosts. In both redshift intervals, the AGN number
density  depends on  the host  mass, peaking  at an  intermediate mass
range and  decreasing toward higher and lower  mass.  Furthermore, the
AGN host mass  function peaks at a higher mass  in the higher redshift
interval.  Our low redshift interval  suffers from low number statistics with on
average $\sim$7 objects per mass bin, compared to $\sim$20 objects per mass bin in
high  redshift  interval.  However,  the  peak  around  a stellar  mass  of
10$^{10.3}$  -   10$^{10.8}$  M$_{\odot}$  is   quite  possibly  real.
For example, \citet{Greene07b} have measured  the local BH mass function  of type 1
SDSS AGNs  at $z$  $<$ 0.3. Although  suffering from  an incompleteness
problem,  there appears to  be a  turnover of  their BH  mass function
around a BH mass of 10$^{7}$ M$_{\odot}$, which corresponds on average
to a host stellar  mass of 10$^{10.2}$ M$_{\odot}$ \citep{Best05}. The
statistical significance  for the low-redshift AGN  host mass function
peaking  in the  second mass  bin  was estimated  using a  Monte-Carlo
simulation.  Basically,  we perturbed the  mass function at  each mass
bin  ten  thousand  times  assuming  a normal  distribution  with  the
1-$\sigma$ deviation  equal to  the measured error.  The probabilities
for the peak at first, second  and third bins are 10\%, 80\% and 10\%,
respectively,  which indicates  that  the low-redshift  AGN host  mass
function most likely peaks in the second mass bin.

To further demonstrate that the AGN host mass function peaks at a higher
mass in  our high redshift interval, the  solid line in the  upper panel of
Fig.~\ref{LF_MHost} shows the SDSS AGN  result at $z$ $<$ 0.3 obtained
by   multiplying   the  fraction   of   galaxies   hosting  AGNs   from
\citet{Kauffmann03a} with the galaxy mass  function at 0.2 $<$ $z$ $<$
0.4 from  \citet{Perez-Gonzalez08}. The galaxy  mass function obtained
by \citet{Perez-Gonzalez08} is based  on a rest-frame near-IR selected
galaxy  sample  and is  generally  consistent  with  results of  other
studies (any  differences are within  0.3 dex (see their  Fig.4)).  At
$z$ $<$  0.7, the result of \citet{Perez-Gonzalez08}  is complete down
to stellar mass of 10$^{9}$  M$_{\odot}$, deep enough for our purpose.
The normalization of the AGN fraction of \citet{Kauffmann03a} has been
decreased by  a factor of 8  to match our  normalization (a discussion
about  the difference  between the  SDSS   and  our low-redshift
samples  can be found  in \S~\ref{red_DAGN}).  With much  higher number
statistics, the  SDSS result  shows the AGN  host mass  function peaks
around 10$^{10.7}$  M$_{\odot}$. We then  carried out a  simulation by
assuming the  0.4 $<$ $z$ $<$  0.7 AGN mass  function actually follows
the SDSS result  at $z$ $<$ 0.3.   For each object at 0.4  $<$ $z$ $<$
0.7, a host mass is assigned randomly with a relative probability that
follows the SDSS  result.  The range of   simulated stellar mass is
from 10$^{9}$  to 10$^{12.5}$ M$_{\odot}$.  We  also assumed that the
total probability  in this  mass range is  equal to 1.   Combining the
simulated stellar mass and the observed $z^{\rm limit}_{\rm xray}$ and
$z^{\rm  limit}_{R}$   measured  in  \S~\ref{Method_NumDen},   we  can
construct the AGN mass function at  0.4 $<$ $z$ $<$ 0.7 using the same
mass bins as for the observed data.  Ten thousand simulations indicate
that the  probability is 99.6\% for  the mass function to  peak in one
of the two lowest mass  bins. That is, the  simulated host mass function
at 0.4 $<$ $z$ $<$ 0.7 (using  the SDSS low-z data as the basis of the
simulated data)  would likely (99.6\%)  peak in the two  low-mass bins
(i.e. $<$  10$^{11}$ M$_{\odot}$), while  the observations at  0.4 $<$
$z$  $<$ 0.7 find  a significant  peak at  higher mass  ($>$ 10$^{11}$
M$_{\odot}$).   This result indicates  that the  tendency for  the AGN
mass  function in  the high  redshift interval  to peak  at higher  mass is
significant (99.6\%).

The simulations shown above test  the pure number statistics. In order
to account for the possible  selection bias that the low-mass galaxies
harbor  faint AGNs  due to  the  mild correlation  between the  galaxy
stellar mass  and the central BH  mass coupled with  a given Eddington
ratio distribution, we first created  a set of stellar masses with the
relative probability  following the galaxy stellar  mass function from
\citet{Perez-Gonzalez08}. These stellar  masses were then converted to
the  BH mass distribution  using the  average ratios  of BH  masses to
galaxy stellar masses  as a function of stellar  masses and associated
errors   from  \citet{Best05}.   For   the  Eddington   ratio  between
[10$^{-5}$,  1], we  assumed a  probability distribution  of $P(L_{\rm
bol}/L_{\rm Edd})$  d($L_{\rm bol}/L_{\rm Edd}$)  $\propto$ 1/($L_{\rm
bol}/L_{\rm Edd}$)$^{n}$, where $n$  is a free parameter.  By assuming
$L_{\rm bol}/L_{\rm  x}=20$, the simulated AGN sample  is then defined
as objects  with $L_{\rm x}$ $>$  10$^{42}$ erg s$^{-1}$  and the host
mass function of  this sample can be constructed.   We simulated about
ten  times for each  $n$ value  which is  given in  a set  of discrete
numbers starting at  zero and increasing in a step  of 0.5.  At $n$=2,
the simulated AGN mass function  shows a similar trend to our measured
mass function  in the redshift  interval of 0.4  $<$ $z$ $<$  0.7. The
peak number density  is about four times larger  than the value around
the lowest  measured mass (10$^{10}$ $M_{\odot}$).   This result shows
that when the Eddington ratio distribution is strongly biased toward a
low value,  the observed turnover of  the AGN mass function  is due to
the selection bias.  However, the corresponding simulated fractions of
galaxies  hosting  AGNs are  an  order  of  magnitude lower  than  the
observed ones. This implies the turnover of the AGN mass function most
likely indicates  a real  decrease of the  number density  of galaxies
with active BHs.

\begin{figure}
\epsscale{1.0}
\plotone{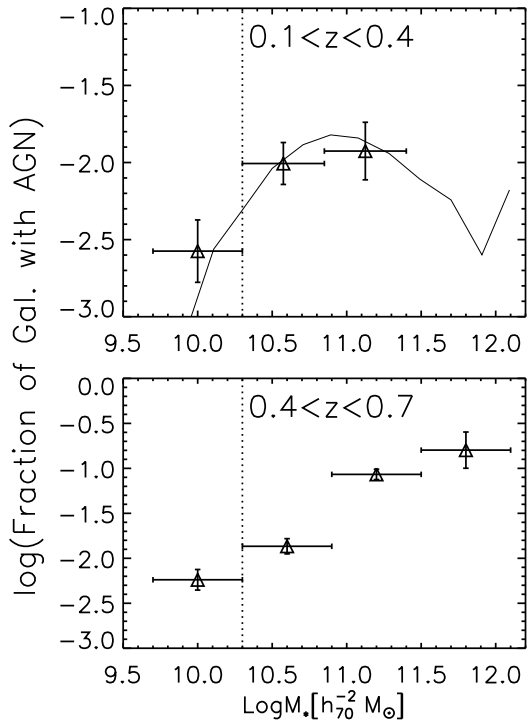}
\caption{  \label{Dwarf_Mass_frac} The  fraction  of galaxies  hosting
active nuclei as  a function of host stellar mass  in the two redshift
intervals of  0.1 $<$  $z$ $<$ 0.4  and 0.4  $<$ $z$ $<$  0.7.  Dotted
lines show the dividing mass between low-mass and massive hosts. The solid
line in the upper panel shows the trend of the SDSS powerful AGN fraction 
with the host stellar mass from \citet{Kauffmann03a} whose normalization has been
decreased by a factor of 8 (a discussion
about  the difference  between the  SDSS   and  our low-redshift
samples  can be found  in \S~\ref{red_DAGN}).}
\end{figure}

Compared to the galaxy mass  function, the unique feature of these two
AGN mass  functions is  that they do  not increase  monotonically with
decreasing mass.  A  physical parameter more related to  the AGN cycle
and BHOF is  the fraction of galaxies hosting  active nuclei, which is
defined  by dividing the  AGN host  mass function  by the  galaxy mass
function    at    a    given     stellar    mass,    as    shown    in
Fig.~\ref{Dwarf_Mass_frac}.  The 0.2  $<$ $z$ $<$ 0.4 and  0.4 $<$ $z$
$<$ 0.6  galaxy mass functions from  \citet{Perez-Gonzalez08} are used
for   our    low   and   high    redshift   intervals,   respectively.
Fig.~\ref{Dwarf_Mass_frac}  shows  that  more  massive  galaxies  host
active  nuclei more  frequently in  both redshift  intervals.   In the
local universe, it has been found that the AGN fraction depends on the
host    stellar   mass   \citep{Kauffmann03a,    Gallo07,   Decarli07,
Sivakoff08},   host    Hubble   type   \citep{Ho97}    and   BH   mass
\citep{Heckman04,  Greene07b}.  Note that  these three  parameters are
roughly  correlated  with  each   other.   In  general,  the  fraction
harboring active  nuclei is lower  for lower-mass local  systems.  Our
result indicates that such a  dependence also exists at high redshift.
However,  there  are differences  in  the  behavior  between low-  and
high-redshift.   \citet{Kauffmann03a}  show their  $z<$  0.3 SDSS  AGN
fraction  increases with  stellar mass  from low-mass  galaxies  up to
galaxies  with 10$^{11}$  M$_{\odot}$  and then  drops quickly  toward
higher mass  galaxies, which is shown  as the solid line  in the upper
panel of  Fig.~\ref{Dwarf_Mass_frac}.  Our result for 0.1  $<$ $z$ $<$
0.4 is consistent  with their work, particularly since  it indicates a
leveling out of the AGN fraction near 10$^{11}$ M$_{\odot}$.  However,
our sample contains too few hosts above this limit to confirm the drop
toward even higher masses.  At 0.4$<$z$<$0.7, our study shows that the
AGN fraction increases  all the way from low-mass  galaxies to massive
galaxies  with $M_{*}$  = 10$^{12}$  $M_{\odot}$.  Note  that  our AGN
fraction for  massive galaxies ($M_{*}$ $>$  10$^{11}$ M$_{\odot}$) is
consistent  with  other  studies   of  X-ray  AGNs  at  this  redshift
\citep{Bundy07,  Alonso-Herrero08}, while  showing that  this  lack of
turn-over  continues up to  higher mass.   The difference  between our
high redshift interval and the SDSS local universe may be explained by
the cosmic ``downsizing''  evolution of MBHs \citep[e.g.][]{Barger05}.
Massive BHs in massive galaxies  have accumulated their masses at high
redshift and become quiescent in the local universe.

Note  that our  AGNs are  defined as  those with  $L_{\rm 2-8keV}^{\rm
rest}$ $>$ 10$^{42}$ erg s$^{-1}$.   It is obvious that the AGN number
density  or  AGN  fraction  depends  on  the  depth  of  surveys,  or,
equivalently, the  limiting Eddington accretion  ratio for a  given BH
mass.   As shown in \S~\ref{DAGN_DEF}, the  limiting Eddington  ratio is
0.02  for our  lowest stellar  mass of  10$^{9.7}$ M$_{\odot}$.  For a
galaxy  with  stellar  mass  of 10$^{11}$  M$_{\odot}$,  the  limiting
Eddington ratio  is 0.001. To have  a sense of the  missed fraction of
AGNs with lower Eddington ratios,  we used the currently most complete
local AGN sample from \citet{Ho95} as a comparison. Our sample of AGNs
should  miss   a  significant  fraction   of  high-ionization  Seyfert
galaxies,  as  local  Seyferts   have  a  median  Eddington  ratio  of
1.3$\times$10$^{-4}$  \citep{Ho08}.  We  may  miss all  low-ionization
nuclear emission-line  regions (LINERs), as  none of the  local LINERs
has an Eddington ratio $>$  0.001 \citep{Ho08}.  Overall, only 5\% and
10\%  of all local  AGNs have  Eddington ratios  larger than  0.01 and
0.001, respectively \citep{Ho08}.  Therefore, the fraction of galaxies
with active BHs  at high redshift may be much  larger than our result,
which  is limited  to $L_{\rm  2-8keV}^{\rm rest}$  $>$  10$^{42}$ erg
s$^{-1}$. Identifying  these additional  AGNs is difficult  because of
the overlap of their X-ray luminosities with those resulting from star
formation  \citep{Ranalli03}.   However,  it  is  still  important  to
understand the fraction of  galaxies hosting AGNs with relatively high
Eddington ratios and their trends  with host masses, as these AGNs may
be in the phase when feedback from MBHs is most important and also the
main stage of BH growth through accretion.

\subsection{X-ray Luminosity Function of AGNs in Low-Mass Hosts}\label{XLF_DAGN}

\begin{figure}
\epsscale{1.0}
\plotone{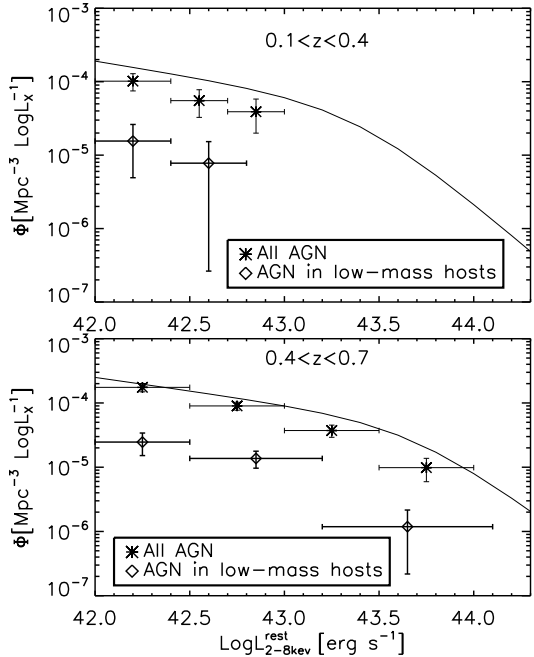}
\caption{  \label{LF_xray}  The rest-frame  2-8  KeV X-ray  luminosity
function of   AGNs in low-mass hosts, and of  all AGNs in  the two redshift  intervals of
0.1$<$ $z$ $<$0.4 and 0.4$<$ $z$  $<$0.7. The two solid lines show the XLF
obtained  by \citet{Barger05}, based on  the CDFN,  CDFS and
CLASXS fields.  }
\end{figure}

Fig.~\ref{LF_xray}  shows the  X-ray  luminosity functions  of AGNs  in
low-mass hosts (diamonds) and all  AGNs (asterisks) in the two redshift
intervals of 0.1 $<$ $z$ $<$0.4 and 0.4$<$ $z$ $<$0.7. Two solid lines
show the  XLFs obtained by  \citet{Barger05}, based on the  CDFN, CDFS
and CLASXS fields. The XLFs of  our AGN host sample with $R<$24 closely
match  their results  for  the  higher redshift  interval.   In the  low
redshift  interval, there is  about a  factor of  two difference  in the
highest  luminosity bin.   This is  most likely  caused by  the cosmic
variance due to the small  comoving volume at low redshift. 

The  total X-ray  energy  density of  AGNs  in low-mass  hosts can  be
measured  by integrating  the XLF  where there  are data  points.  The
measurements       give       (2.2$\pm$1.4)$\times$10$^{37}$       and
(1.4$\pm$0.4)$\times$10$^{38}$ erg s$^{-1}$  Mpc$^{-3}$ at 0.1 $<$ $z$
$<$ 0.4  and 0.4 $<$ $z$  $<$ 0.7, respectively.  Similarly, the X-ray
energy density of all AGNs  can be obtained.  The contribution of AGNs
in low-mass  hosts to the  total X-ray energy density  is (11$\pm$8)\%
and  (14$\pm$5)\%   in  the  lower  and   higher  redshift  intervals,
respectively.   Therefore, these AGNs  are energetically  important to
the cosmic X-ray background.

\subsection{Redshift Evolution of  AGNs in Low-Mass Hosts}\label{red_DAGN}

\begin{figure}
\epsscale{1.0}
\plotone{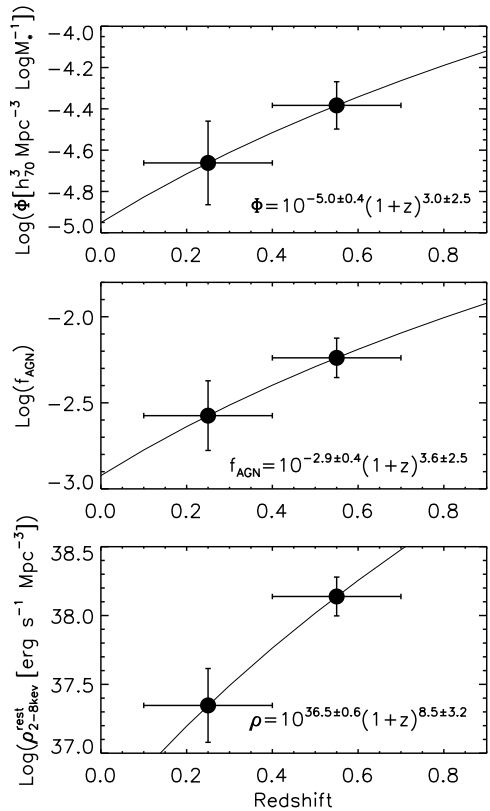}
\caption{ \label{DAGN_red}  The redshift  evolution of   AGNs with
 host  stellar mass 9.7  $<$ log$M_{*}$/M$_{\odot}$ $<$  10.3. From
top  to bottom:  the comoving  number density;  the fraction  of low-mass
galaxies hosting active nuclei;  and the comoving X-ray energy density.}
\end{figure}

As  shown in Fig.~\ref{DAGN_red},  AGNs in  low-mass hosts  show strong
redshift  evolution.   The  number  density  increases  with  redshift
following      $\Phi_{\rm       DAGN}$      =      10$^{-5.0{\pm}0.4}$
(1+$z$)$^{3.0{\pm}2.5}$    Mpc$^{-3}$   log(M$_{\odot}$)$^{-1}$,   the
fraction  of low-mass  galaxies with  AGNs evolves  as $f_{\rm  AGN}$ =
$10^{-2.9{\pm}0.4}(1+z)^{3.6{\pm}2.5}$ and  the energy density follows
$\rho_{\rm  DAGN}$  =  10$^{36.5{\pm}0.6}$(1+$z$)$^{8.5{\pm}3.2}$  erg
s$^{-1}$ Mpc$^{-3}$ in the  redshift range of 0$<$$z$$<$0.7. Note that
these three quantities are related to each other to some degree.

To  evaluate our  result  for  cosmic evolution  of  AGNs in  low-mass
galaxies,  we compared  it  at  low-redshift to  those  based on  SDSS
optical-emission-line-selected AGNs.   \citet{Greene07b} show that the
spatial  number  density  of  BHs between  10$^{6.5}$  and  10$^{7.5}$
M$_{\odot}$ hosted by SDSS  narrow-line AGNs from \citet{Heckman04} is
around 6$\times$10$^{-5}$  Mpc$^{-3}$ logM$_{\rm BH}^{-1}$  (see their
Fig.10). Assuming the  host galaxies of these BHs  have $M_{*}$ $\sim$
10$^{10} $ M$_{\odot}$, this number density is three times higher than
the  prediction   of  $\sim$(2.0$\pm$1.5)$\times$10$^{-5}$  Mpc$^{-3}$
logM$_{*}^{-1}$  at a  redshift of  0.15 based  on the  number density
evolution of our AGNs in low-mass hosts. Note that our result has been
corrected for broad-line AGN.   However, the correction is quite small
($\sim$10\%) as shown in \S~\ref{VmaxMethod} and Fig.~\ref{BLAGNFRAC}.
The difference  could be partly  caused by our low  number statistics.
Although the limiting Eddington ratios of the \citet{Greene07b}'s AGNs
are similar to ours  ($\sim$ 0.01), the different selection algorithms
(optical  emission diagnostics  vs.   X-ray emission)  may be  another
reason  for the  discrepancy.  The  optical emission  diagnostics miss
AGNs when the contrast of narrow line emission to host galaxy light is
low.   At low redshift,  the slit  for the  spectroscopic observations
covers  only part  of  the host  galaxies  so that  host galaxy  light
dilution is  not quite so severe.  \citet{Kauffmann03a}  show that the
SDSS  optical emission  line  diagnostics can  recover  the AGNs  with
Eddington  ratios below  0.01 for  host galaxies  with  $M_{*}$ $\sim$
10$^{10}  $  M$_{\odot}$,  if  the  emission lines  are  not  strongly
extincted. The X-ray emission  method misses AGNs whose X-ray emission
is obscured  so that their  observed X-ray luminosity falls  below our
threshold for  AGN definition.  It is obvious  that Compton-thick AGNs
($N_{H}$ $>$ 10$^{24}$ cm$^{-2}$) will be missed in the 2-10 keV X-ray
survey. However,  if AGNs in low-mass hosts  have intrinsically weaker
X-ray   emission   compared   to   more   massive   host   AGNs,   the
heavily-extincted  but  Compton-thin  AGNs  (10$^{23}$  cm$^{-2}$  $<$
$N_{H}$  $<$ 10$^{24}$  cm$^{-2}$)  may still  be  missed.  Our  X-ray
spectral fits of AGNs in low-mass hosts with X-ray counts $>$ 100 (Shi
et al.  2008, in preparation) have shown that only $\sim$10\% of these
AGNs have $N_{\rm HI}$  $>$ 10$^{23}$ cm$^{-2}$. \citet{Heckman05} did
show that the emission-line-selected  AGN sample is more complete than
the X-ray-selected one  at low redshift and that  the SDSS AGN spatial
density  is  about  three   times  higher  than  X-ray-selected  AGNs.
Although at  low redshift optical  emission line diagnostics  are more
complete, X-ray selection identifies more AGNs at high redshift. X-ray
selection also  provides more  uniform selection criteria  (e.g., less
affected  by dilution  by the  host galaxy  extended emission).   As a
summary, the prediction at low-redshift  based on our result of cosmic
evolution of  AGNs in low-mass  galaxies is generally  consistent with
the SDSS result,  with an offset caused by  different selection biases
between           X-ray-emission-selected           AGNs           and
optical-emission-line-selected ones.

%This results in
%underestimating our total energy density  by a factor of 2 compared to
%the integration  over the  fitted analytical XLF  \citep{Barger05}. 

\section{Discussion: Black Hole Occupation Fraction in Low-Mass Galaxies}\label{Discussion}

\subsection{Fraction of Galaxies Hosting AGNs}\label{DIS_FRACAGN}

\S~\ref{COMOVING_ND}  shows that  the AGN  mass function  peaks  at an
intermediate mass  and decreases toward the low-mass regime, in contrast
to  the  monotonic   increase    of   the    galaxy    mass   function.
\S~\ref{COMOVING_ND} also  indicates that the fraction  of galaxies hosting
active   nuclei   decreases   with     decreasing   host   stellar
mass.  Theoretically, the  fraction  of galaxies  hosting  AGNs can  be
estimated by

\begin{equation}
f_{\rm AGN}=f_{\rm BHOF}{\gamma}t_{\rm AGN}
\end{equation} 
where $f_{\rm BHOF}$ is the  black hole occupation function, i.e., the
fraction of  galaxies hosting MBHs  at their centers, $\gamma$  is the
fractional AGN trigger rate, i.e., the fraction of MBHs becomes active
per unit time, and $t_{\rm AGN}$ is the duration of a nuclear activity
episode  with  $L_{X}$$>$10$^{42}$ erg  s$^{-1}$.   The  trend of  AGN
fractions with host mass can be  caused by the host mass dependence of
any one  of the  three factors, $f_{\rm  BHOF}$, $\gamma$  and $t_{\rm
AGN}$.  Different models for MBH  seed formation in the early universe
predict    different    mass    dependences    of    $f_{\rm    BHOF}$
\citep{Volonteri08}.  $t_{\rm AGN}$ is  most likely mass dependent, as
the lower mass systems require  larger Eddington ratios to be brighter
than $L_{X}$$=$10$^{42}$ erg s$^{-1}$.

To get  some idea of  $f_{\rm BHOF}$ of  low-mass galaxies, we  make a
simple  assumption that  each major  merger triggers  one-time nuclear
activity  and  thus  $\gamma$   =  the  merging  rate.   Although  the
measurement of  the merging rate  is subject to  various uncertainties
\citep[e.g.][]{Cassata05, Shi06, Lotz08},  it should be around 0.1-0.2
Gyr$^{-1}$  for  massive galaxies.   Simply  assuming  $\gamma$ =  0.1
Gyr$^{-1}$ for low-mass galaxies, we have :

\begin{equation}
f_{\rm BHOF} = \frac{\rm 0.03 Gyr}{t_{\rm  AGN}}\frac{0.1}{\gamma}; (0.1<z<0.4)
\end{equation} 
\begin{equation}
f_{\rm BHOF} = \frac{\rm 0.06 Gyr}{t_{\rm  AGN}}\frac{0.1}{\gamma}; (0.4<z<0.7)
\end{equation}

In  this simplified  case, as  long as  the duration  of  one episodic
nuclear active  phase is  not long  ($<$0.06 Gyr),  {\it all}
low-mass galaxies with 9.8 $<$  log$M_{*}$ $<$ 10.3 at $z$ $<$ $\sim$1
should harbor BHs at their centers.

\subsection{Accreted BH Mass by AGNs in Low-Mass Galaxies}

The amount of    BH mass accreted by  AGNs in low-mass galaxies since $z$ =  1 can be
measured  by integrating  the  cosmic evolution  of  the X-ray  energy
density of  these AGNs obtained in \S~\ref{red_DAGN}.  This mass should
provide  a  stringent   lower  limit  to  the  BHOF   in  local  low-mass
galaxies. An average X-ray  to bolometric luminosity correction of 19
is  obtained by  using  $L_{\rm bol}/L_{2-10  \rm keV}=17(L_{2-10  \rm
keV}/10^{43}$  erg  s$^{-1})^{0.43}$  \citep{Shankar04},  $L_{\rm  2-8
keV}/L_{2-10  \rm   keV}$=0.86  and  the  mean   X-ray  luminosity  of
7$\times$10$^{42}$    erg    s$^{-1}$.    Assuming    the
mass-to-radiation  conversion efficiency $\varepsilon$=0.1,  the total
accreted   black  hole   mass   in   galaxies   with  9.7   $<$
log($M_{*}$/$M_{\odot}$)    $<$    10.3   since    $z$    $=$   1    is
3.9($\pm$0.9)$\times$10$^{3}$ M$_{\odot}$ Mpc$^{-3}$.

This accreted BH mass must  be hosted in local low-mass galaxies.  The
corresponding total  galaxy mass can  be estimated using  the bulge-BH
relation and bulge-to-disk  ratio.  The lower limit to  the local BHOF
is then the  ratio of the total mass of  galaxies hosting the accreted
BHs to the total local mass  in low-mass galaxies. In practice, we can
alternatively assume all local  low-mass galaxies host BHs.  The lower
limit of  the BHOF is then  the ratio of  the accreted BH mass  to the
assumed total BH mass hosted by all local low-mass galaxies.

The  BH mass  function  can  be determined  from  the galaxy  velocity
dispersion  function  using  the  relationship  between  the  velocity
dispersion and BH mass  ($M_{BH}$-$\sigma$). However, it  is  relatively 
difficult  to  measure  the velocity  dispersion,  resulting  in  incompleteness 
of  the  BH  mass function  at the  low mass  end. Alternatively, it can be
derived  from the  bulge  luminosity function  using the  relationship
between   the   bulge   luminosity    and   the   BH   mass   ($L_{\rm
Bulge}$-$M_{BH}$).  In this case, the bulge-to-disk ratios of galaxies
with  different   morphologies  are  required  to   derive  the  bulge
luminosity  function  from   the  galaxy  total  luminosity  function.
In practice, the two methods  are actually employed at the same time 
to complement each other.  Different studies
produce a  relatively consistent  result for the  total local  BH mass
density, which is around  4.5 $\times$ 10$^{5}$ M$_{\odot}$ Mpc$^{-3}$
with   uncertainty  $<$  2.0$\times$10$^{5}$   M$_{\odot}$  Mpc$^{-3}$
\citep{Aller02,  Shankar04, Marconi04}.   Therefore,  the accreted  BH
mass by AGNs in low-mass hosts since  $z$ $=$ 1 is only a small fraction
(0.8\%) of the total local BH mass density.

To estimate the BHOF in  local low-mass galaxies using the accreted BH
mass during  their AGN  phase since $z$  = 1,  we first  need to
determine the  local successor of  our low-mass AGN hosts.   Given the
cosmic  evolution  of  the  stellar  mass  density  of  galaxies  with
different stellar masses provided by \citet{Perez-Gonzalez08}, the 9.7
$<$ log($M_{*}$/$M_{\odot}$)  $<$ 10.3 interval at  median redshift of
$z$ = 0.5 corresponds  roughly to 9.85$<$ log($M_{*}$/$M_{\odot}$) $<$
10.45 at $z$ = 0.  To estimate the total BH mass associated with local
galaxies with 9.85 $<$ log($M_{*}$/$M_{\odot}$) $<$ 10.45, we measured
the  BH   mass  hosted  by  the  early-type   and  late-type  galaxies
separately,  as  they have  different  bulge  to  total light  ratios,
$f^{\rm  early}_{\rm bulge}$  = 0.85$\pm$0.05  and  $f^{\rm late}_{\rm
bulge}$    =   0.30$\pm$0.05    \citep{Shankar04}.     Assuming   that
log($M_{BH}$/M$_{\odot}$) = (8.20$\pm$0.10)+(1.12$\pm$0.06)log($M_{\rm
bulge}$/10$^{11}$M$_{\odot}$)  \citep{Haring04}, early  and  late type
galaxies with  9.85 $<$ log($M_{*}$/M$_{\odot}$) $<$  10.45 harbor BHs
with  6.85   $<$  log($M_{BH}/M_{\odot}$)   $<$  7.50  and   6.15  $<$
log($M_{BH}/M_{\odot}$) $<$  6.80, respectively.  By  fitting Fig.5 of
\citet{Shankar04} with the formula
\begin{equation}
\Phi=\Phi_{BH}^{*}(\frac{M_{BH}}{M_{BH}^{*}})^{\alpha+1}{\rm exp}[-(\frac{M_{BH}}{M_{BH*}})^{\beta}],
\end{equation}        
we        obtained       $\Phi_{BH}^{*}        =
5.2(\pm0.2){\times}10^{-3}$       Mpc$^{-3}$,       $M_{BH}^{*}      =
8.8(\pm2.0){\times}10^{6}$  M$_{\odot}$,  $\alpha$ =  -0.47($\pm$0.03)
and   $\beta$   =  0.39($\pm$0.01)   for   early-type  galaxies,   and
$\Phi_{BH}^{*} = 1.9(\pm0.2){\times}10^{-2}$ Mpc$^{-3}$, $M_{BH}^{*} =
3.3(\pm1.7){\times}10^{5}$  M$_{\odot}$,  $\alpha$ =  -0.57($\pm$0.09)
and $\beta$ = 0.34($\pm$0.02) for late-type galaxies.  Integrating the
above equation  over the range  of interest, the local early-  and late-type
galaxies  with 9.85$<$ log($M_{*}$/$M_{\odot}$)  $<$ 10.45  harbor 
   total BH   masses    of  2.1$\times10^{4}$     and
1.2$\times10^{4}$  M$_{\odot}$ Mpc$^{-3}$, respectively. Therefore,  the accreted
BH mass by  AGNs in low-mass hosts from $z$  $=$ 1 contributes 12\% of  the local BH
mass  hosted  in galaxies  with  9.85$<$ log($M_{*}$/$M_{\odot}$)  $<$
10.45.

Three main  uncertainties associated with this  percentage include the
accreted BH mass,  the stellar mass of local  counterparts of our high
redshift  low-mass hosts  and the  local BH  mass function.   As shown
above the measurement  error of the accreted BH mass  is about 20\%. A
larger uncertainty  for this  accreted BH mass  is from  the uncertain
estimate of AGN  accretion in low-mass galaxies at 0.7  $<$ $z$ $<$ 1.
The current assumption is that it follows the trend at $z$ $<$ 0.7. If
the energy density of AGNs in the low-mass host is flat at $z$ $>$ 0.7,
the  accreted BH  mass  will decrease  by  about a  factor  of 2.   To
estimate the second  uncertainty, we can assume an  extreme case where
all our AGNs in low-mass hosts are  at $z$ = 1. In this case, the local
counterparts  have 10.1$<$ log($M_{*}$/$M_{\odot}$)  $<$ 10.7  and the
accreted BH mass contribution is 6\%.  The uncertainty in the local BH
mass function  is about a factor  of 3, from  comparing different studies
\citep{Aller02,   Shankar04,   Marconi04}.    Therefore,   the   final
uncertainty  could  be  a  factor  of  4, estimated  by  adding  the  above  errors
quadratically.

A  more important  factor, resulting  in severe  underestimate  of the
accreted BH mass by AGNs,  is the X-ray obscuration correction for the
observed   X-ray  flux   plus   the  significant   number  of   missed
heavily-obscured AGNs in current  X-ray surveys.  This problem is even
worse for  AGNs in  low-mass hosts,  as (1) they  on average  have low
intrinsic X-ray  luminosities, with  many falling below  our selection
criteria of $L_{X}$$>$10$^{42}$ erg  s$^{-1}$ and (2) the structure of
the circumnuclear material may vary  with the luminosity in a way that
causes lower luminosity objects to be more obscured, as implied by the
decrease in the fraction of  type 2 objects with increasing luminosity
\citep[e.g.][]{Barger05}.     We     calculated    the    degree    of
under-estimation in two ways. First,  we measured the accreted BH mass
of 3.4$\times10^{4}$ M$_{\odot}$ Mpc$^{-3}$ by all X-ray-detected AGNs
from $z$ = 1 by integrating  the X-ray energy density evolution of all
AGNs  (see \S~\ref{XLF_DAGN}).   \citet{Shankar04}  obtained total  BH
mass of  $\sim$1.3$\times$10$^{5}$ M$_{\odot}$ Mpc$^{-3}$  accreted by
all AGNs from $z$ = 1 with a correction for obscuration and accounting
for missing Compton-thick objects.  Therefore, the accreted BH mass by
AGNs  in low-mass  galaxies is  underestimated by  a factor  of  4, by
assuming that the obscuration of these  AGNs is similar to that of all
AGNs. The second  way is to compare the distribution  of the HI column
density of observed AGNs to the intrinsic distribution to estimate the
fraction of missing AGNs.  Our X-ray spectral fits of AGNs in low-mass
galaxies with X-ray counts $>$  100 (Shi et al.  2008, in preparation)
have shown that  only 10\% of such AGNs have  HI column density $N_{\rm
HI}$  $>$ 10$^{23}$ cm$^{-2}$.   The study  of local  Seyfert galaxies
shows that 75\% of Seyfert  2 galaxies have $N_{\rm HI}$ $>$ 10$^{23}$
cm$^{-2}$  \citep{Risaliti99}.  If  AGNs  in  low-mass  galaxies  have
structures  of  circumnuclear material  similar  to  those in  Seyfert
galaxies,  75\% of  the low-mass  galaxies  should be  missed and  the
accreted BH mass by all such  AGNs is underestimated by a factor of 4,
similar to the value estimated by the first method.

Therefore,  the  accreted  BH  mass  by all  AGNs  in  low-mass  hosts
including  missed  heavily-obscured  ones  from $z$  $=$  1,  probably
contribute $\sim$  50\% of the local  BH mass hosted  in galaxies with
9.85$<$  log($M_{*}$/$M_{\odot}$)  $<$  10.45.   As discussed  at  the
beginning of this section, this percentage is obtained by assuming all
such galaxies  harbor BHs at their  centers. A percentage  of 50\% can
also be interpreted  as the lower limit of the  BHOF in local galaxies
with 9.85 $<$ log($M_{*}$/$M_{\odot}$) $<$ 10.45, i.e., at least 1 out
of 2  local galaxies with 9.85 $<$  log($M_{*}$/$M_{\odot}$) $<$ 10.45
harbor  BHs.   Accretion  at  higher  redshift ($z$  $>$  1),  at  low
Eddington ratios, or the mass  growth through the BH merging processes
may provide the remaining BH mass.

The above results strongly favor a scenario where the nuclear activity
is  a  necessary ingredient  during  low-mass  galaxy evolution.   The
fraction   of   low-mass   galaxies   hosting   active   nuclei   (see
\S~\ref{DIS_FRACAGN}) and  the amount of the accreted  black hole mass
in low-mass  host AGNs (see  this section) implies that  a significant
fraction of low-mass galaxies in the local universe harbor black holes
at their centers and these black hole masses are assembled through the
active accretion phase ($L_{X}>10^{42}$ erg s$^{-1}$) at $z$
$<$ 1.0.

\section{Conclusions}\label{Conclusions}

We have studied a sample of  X-ray AGNs in low-mass host galaxies with
stellar  mass   of  5$\times$10$^{9}$  M$_{\odot}$   $<$  $M_{*}$  $<$
2$\times$10$^{10}$   M$_{\odot}$  out  to   $z$  $\sim$1.    Our  main
conclusions are:

%(1) The  bolometric luminosity  per stellar  mass in  dwarf AGN  is as
%strong  as in  massive ones  and can  easily exceed  supernova energy,
%implying  the AGN  energy feedback  may affect  the dwarf  host galaxy
%evolution.

(1) By  including  AGNs  in   more  massive  host  galaxies,  we  have
constructed the  stellar mass function of AGN  host galaxies extending
down to the  low-mass regime in two redshift intervals  of 0.1 $<$ $z$
$<$ 0.4 and 0.4  $<$ $z$ $<$ 0.7.  We have found  the AGN host stellar
mass  function peaks  at an  intermediate  mass range,  with the  peak
shifting toward higher mass at higher redshift.

(2) By comparing to AGNs in more massive hosts, we have found that the
fraction  of galaxies hosting  active nuclei  depends strongly  on the
host mass, increasing from  low-mass galaxies continuously to the most
massive ones.  The fraction of low-mass galaxies hosting active nuclei
suggests that  a large fraction  of such galaxies  at 0 $<$ $z$  $<$ 1
harbor black holes  at their centers as long as  the low-mass host AGN
lifetime  with $L_{X}$  $>$ 10$^{42}$  erg s$^{-1}$  is not
long ($<$ 0.06 Gyr).

(3) The X-ray luminosity functions of AGNs in low-mass hosts have been
constructed in two  redshift intervals of 0.1 $<$ $z$  $<$ 0.4 and 0.4
$<$ $z$ $<$ 0.7.  AGNs  in low-mass hosts contribute $\sim$10\% of the
X-ray energy  density of all the AGNs  in 0 $<$ $z$  $<$ 1, indicating
that such  AGNs make an energetically significant  contribution to the
cosmic X-ray background.

(4) AGNs  in low-mass hosts  show strong  redshift evolution  in their
comoving  number density, the  fraction of  such galaxies  with active
nuclei and the comoving  radiation energy density.  By integrating the
X-ray luminosity function  of these AGNs over the  redshift range of 0
$<$  $z$  $<$  1,  the  accreted  black hole  mass  in  galaxies  with
5$\times$10$^{9}$  M$_{\odot}$   $<$  $M_{*}$  $<$  2$\times$10$^{10}$
M$_{\odot}$ is  (3.9$\pm$0.9)$\times$ 10$^{3}$ M$_{\odot}$ Mpc$^{-3}$.
This number  gives a  strong lower  limit of 12\%  to the  fraction of
local  low-mass galaxies  harboring  black holes,  which  may be  much
higher ( $>$ 50\%) if the dusty torus of the AGNs in low-mass galaxies
has  similar structure  to or  is  more opaque  than that  of AGNs  in
massive host galaxies.

\acknowledgements
We thank  the anonymous referee  for detailed comments. Support for  this work was
provided by  NASA through contract  1255094 issued by  JPL/ California
Institute of Technology.

\clearpage

\begin{deluxetable}{lcccccccccccccccc}
\tabletypesize{\scriptsize}
\tablecolumns{7}
\tablecaption{\label{xrayfields} X-ray Fields}
\tablewidth{0pt}
\tablehead{ 
\colhead{FIELD}                      &     \colhead{Area}                   & \colhead{$F^{\rm limit}_{\rm Xray}$} & 
\colhead{Ref}                        &     \colhead{Optical photometry}     & \colhead{Ref}                        & 
\colhead{Secure redshift}            &     \colhead{Ref} \\
\colhead{     }                      &     \colhead{deg$^{2}$}              & \colhead{erg s$^{-1}$ cm$^{-2}$}     & 
\colhead{}                           &     \colhead{}                       & \\
\colhead{ (1) }                      &     \colhead{  (2)    }              & \colhead{    (3)   }                 & 
\colhead{ (4) }                      &     \colhead{  (5)    }              & \colhead{    (6)   }                 &
\colhead{    (7)   }                 &     \colhead{  (8)    }           }
\startdata
AEGIS           & 0.67  &  5.0$\times$10$^{-16}$        & 1    & B, R, I, K$_{s}$     & 2,3         & zquality=3, 4                   & 4       \\
CDF-N           & 0.12  &  1.4$\times$10$^{-16}$        & 5    & U, B, V, R, I, z, HK & 6           & all $z$ without label $'$s$'$   &6, 7    \\
CDF-S           & 0.11  &  2.8$\times$10$^{-16}$        & 8    & U, B, V, R, I, 914nm & 9           & qual=2                          &10      \\
CLASXS          & 0.4   &  3.0$\times$10$^{-15}$        & 11   & B, V, R, I, $z^{'}$  & 12          & all listed $z$ are secure       &12      \\
XMMLSS          & 1     &  5.8$\times$10$^{-15}$        & 13   & B, V, R, I, J, K     & 14,15,16,17 & q$_{-}$z=3, 4, 13, 14, 23, 24   &18       \\
\enddata
\tablecomments{  Col.(1): The field name. Col.(2): The area of the field size in square degree. Col.(3): The limiting
X-ray flux in 2-8 keV. Col.(4): The reference for the X-ray data. Col.(5): The available optical/near-IR photometry. Col.(6):
The reference for the optical/near-IR photometry. Col.(7): The definition of the secure spectroscopic redshift in each field.
Col.(8): The reference for the spectroscopic redshift.\\
References -- (1) \citet{Nandra05,  Laird08}; (2) \citet{Coil04}; (3) \citet{Bundy06};
(4) \citet{Davis03,         Davis07};         (5) \citet{Alexander03};
(6) \citet{Barger03};   (7) \citet{Barger02};  (8) \citet{Alexander03};
(9) \citet{Wolf04}   ;   (10) \citet{Szokoly04};   (11) \citet{Yang04};
(12) \citet{Steffen04};                      (13) \citet{Chiappetti05};
(14) \citet{LeFevre04};                       (15) \citet{McCracken03};
(16) \citet{Radovich04}; (17) \citet{Iovino05}; (18) \citet{LeFevre05} }
\end{deluxetable}

\begin{deluxetable}{lcccccccccccccccc}
\tabletypesize{\scriptsize}
\tablecolumns{7}
\tablecaption{\label{xrayfield_LF} The number of X-ray objects in all fields}
\tablewidth{0pt}
\tablehead{ 
\colhead{FIELD}         &     \colhead{CDF-N} &  \colhead{CDF-S}  &  \colhead{CLASXS}  & \colhead{AEGIS}  & \colhead{XMMLSS}
}
\startdata
Total                   &     503            &  326             &   525              &   1318      & 286  \\
Spec-Observed           &     439            &  210             &   422              &   357       & 23   \\        
\enddata
\end{deluxetable}

\begin{deluxetable}{llllllllllllllll}
\tabletypesize{\scriptsize}
\tablecolumns{7}
\tablecaption{\label{BC_model} The Parameters of Stellar Synthesis Models.}
\tablewidth{0pt}
\tablehead{ 
\colhead{parameters}           &     value  \\
}
\startdata
Simple stellar populations                                     &  \citet{Chabrier03} IMF and Padova 1994 evolutionary tracks\\
Metallicity                                                    &  0.0001, 0.0004, 0.004, 0.008, 0.02 ($Z_{\odot}$), 0.05 \\
visual extinction $\tau_{v}$                                   &  [0.0, 4.0] with a step of 0.5 in logarithm \\
e-folding time $\tau$ for   exponential star-formation history &  [0.05, 8.91] Gyr with a step of 0.25 in logarithm, 100 Gyr       \\
fraction of ejected gas to be recycled $\epsilon$              & 0.001, 0.01, 0.1, 1\\
galaxy age                                                     &  [0.001, 15.85] Gyr with a step of 0.1 in logarithm \\
\enddata
\end{deluxetable}

\clearpage

         \LongTables % optionally
         \begin{landscape}
\begin{deluxetable}{lllllcll}
\tabletypesize{\scriptsize}
\tablecolumns{7}
%\rotate
\tablecaption{\label{target} Sample of AGNs in Low-Mass Host Galaxies}
\tablewidth{0pt}
\tablehead{ \colhead{source}             &   \colhead{$z$}          &  \colhead{RA$_{\rm X}$, DEC$_{\rm X}$} & 
            \colhead{ $D_{\rm opt}$ }    &   \colhead{Phtometry}   &  \colhead{$f_{\rm X}$}                &
            \colhead{ log(Mass) }        &   \colhead{ $f_{\rm X}/f_{\rm r}$ } \\
            \colhead{      }             &   \colhead{     }        &  \colhead{   }                        & 
            \colhead{[$''$]}             &   \colhead{[in AB system]} &  \colhead{ [10$^{-15}$ergs$^{-1}$cm$^{-2}$] }                      &
            \colhead{[log(M$_{\odot}$)]}&   \colhead{     }  \\
            \colhead{ (1)  }             &   \colhead{ (2) }        &  \colhead{(3)}                        & 
            \colhead{ (4)  }             &   \colhead{ (5) }        &  \colhead{ (6) }                      &
            \colhead{ (7)  }             &   \colhead{ (8) }  
}
\startdata

   AEGIS123  &    0.92 &   14 15 39.2  +52 08 49.6 &  0.98 &                              24.35, 23.95, 23.71,  --   &   2.88 &    9.52$^{+0.40}_{-0.25}$ &  1.46 \\
   AEGIS232  &    0.81 &   14 16 16.8  +52 21 35.6 &  1.37 &                              24.00, 23.31, 22.60, 21.62 &   1.06 &   10.05$^{+0.08}_{-0.07}$ &  0.32 \\
   AEGIS375  &    0.48 &   14 17 24.6  +52 30 25.0 &  0.31 &                              20.23, 19.99, 19.66, 18.51 &   19.1 &   10.21$^{+0.07}_{-0.03}$ &  0.33 \\
   AEGIS555  &    0.84 &   14 18 25.2  +52 49 20.8 &  0.77 &                              24.48, 24.06, 23.38,  --   &   2.14 &    9.64$^{+0.15}_{-0.13}$ &  1.26 \\
   AEGIS655  &    0.71 &   14 19 06.5  +52 38 55.8 &  1.29 &                              23.52, 22.65, 22.18, 20.93 &   3.26 &   10.09$^{+0.04}_{-0.08}$ &  0.56 \\
   AEGIS723  &    0.66 &   14 19 30.8  +52 56 17.3 &  0.42 &                              25.99, 23.74, 22.76, 21.76 &   5.26 &   10.25$^{+0.35}_{-0.19}$ &  2.54 \\
   AEGIS795  &    0.46 &   14 20 01.4  +52 53 10.7 &  0.96 &                              23.85, 21.91, 21.32, 20.16 &   2.40 &   10.22$^{+0.08}_{-0.01}$ &  0.25 \\
   AEGIS969  &    0.91 &   14 20 59.8  +52 56 04.3 &  0.92 &                              23.06, 22.71, 22.00, 20.62 &   2.28 &   10.06$^{+0.00}_{-0.00}$ &  0.37 \\
   AEGIS1008 &    0.43 &   14 21 15.9  +53 19 48.6 &  0.38 &                              22.27, 21.23, 20.81, 20.15 &   3.15 &    9.92$^{+0.31}_{-0.01}$ &  0.18 \\
   AEGIS1159 &    1.24 &   14 22 14.8  +53 23 54.3 &  1.90 &                              24.02, 23.94, 23.35,  --   &   1.55 &   10.12$^{+0.30}_{-0.25}$ &  0.67 \\
   AEGIS1303 &    1.00 &   14 23 26.1  +53 30 03.7 &  0.79 &                              22.54, 21.89, 21.64, 20.69 &   11.4 &   10.23$^{+0.06}_{-0.01}$ &  0.84 \\
   CDF-N5    &    0.56 &   12 35 21.3  +62 16 28.1 &  0.05 &         23.20, 22.82, 22.42, 21.83, 21.35, 21.20, 20.59 &   4.22 &   10.20$^{+0.23}_{-0.33}$ &  0.37 \\
   CDF-N83   &    0.46 &   12 36 08.2  +62 15 53.1 &  1.52 &         23.50, 23.22, 22.62, 21.93, 21.55, 21.50, 20.89 &   3.52 &   10.02$^{+0.11}_{-0.11}$ &  0.37 \\
   CDF-N191  &    0.56 &   12 36 35.9  +62 07 07.7 &  1.43 &         25.40, 25.52, 25.22, 24.43, 23.85, 23.60, 21.59 &   5.31 &    9.93$^{+0.00}_{-0.00}$ &  5.15 \\
   CDF-N194  &    0.56 &   12 36 36.7  +62 11 56.0 &  0.34 &         24.30, 23.42, 23.12, 22.43, 21.85, 21.60, 20.80 &   2.01 &   10.12$^{+0.36}_{-0.26}$ &  0.31 \\
   CDF-N267  &    0.40 &   12 36 51.7  +62 12 21.4 &  0.97 &         24.20, 23.52, 22.62, 22.03, 21.65, 21.40, 20.59 &   2.65 &   10.11$^{+0.10}_{-0.13}$ &  0.31 \\
   CDF-N301  &    0.29 &   12 36 58.7  +62 04 02.4 &  0.62 &         22.20, 21.52, 21.62, 21.63, 21.65, 21.40, 21.09 &   5.69 &    8.66$^{+0.07}_{-0.15}$ &  0.51 \\
   CDF-N441  &    0.63 &   12 37 36.0  +62 18 05.9 &  0.00 &         25.20, 25.02, 24.52, 23.93, 23.45, 23.20, 21.59 &   1.12 &   10.22$^{+0.01}_{-0.05}$ &  0.65 \\
   CDF-S241  &    0.68 &   03 32 39.1  -27 44 39.1 &  1.62 &                25.84, 25.32, 25.44, 25.35, 24.54, 24.39 &   2.14 &    8.35$^{+0.37}_{-0.28}$ &  4.49 \\
   CLASXS42  &    0.49 &   10 31 54.9  +57 45 20.9 &  0.69 &                       24.90, 24.70, 24.10, 24.20, 23.80 &   13.0 &    8.63$^{+0.14}_{-0.11}$ &  9.77 \\
   CLASXS131 &    0.39 &   10 32 42.6  +57 56 20.8 &  0.72 &                       23.30, 23.90, 23.50, 23.10, 23.00 &   5.30 &    8.89$^{+0.16}_{-0.17}$ &  2.45 \\
   CLASXS205 &    0.68 &   10 33 18.1  +57 26 01.4 &  0.65 &                       21.50, 22.00, 21.70, 21.40, 21.10 &   22.0 &    9.43$^{+0.02}_{-0.29}$ &  1.60 \\
   CLASXS231 &    1.38 &   10 33 29.2  +57 47 08.1 &  0.56 &                       23.40, 23.60, 23.20, 23.30, 23.30 &   5.90 &   10.15$^{+0.24}_{-0.31}$ &  1.21 \\
   CLASXS243 &    0.32 &   10 33 34.1  +57 56 01.9 &  0.18 &                       21.90, 21.20, 20.50, 20.20, 19.80 &   8.20 &   10.08$^{+0.20}_{-0.32}$ &  0.25 \\
   CLASXS286 &    0.29 &   10 33 53.2  +57 32 41.0 &  0.48 &                       21.00, 20.90, 20.20, 20.00, 19.60 &   14.0 &    9.61$^{+0.06}_{-0.29}$ &  0.33 \\
   CLASXS322 &    0.20 &   10 34 06.6  +57 56 07.3 &  0.26 &                       21.50, 21.00, 20.50, 20.20, 19.90 &   19.0 &    9.59$^{+0.19}_{-0.53}$ &  0.64 \\
   CLASXS329 &    0.37 &   10 34 09.5  +57 29 53.8 &  0.30 &                       22.60, 21.80, 21.20, 21.00, 20.60 &   5.40 &   10.19$^{+0.12}_{-0.11}$ &  0.31 \\
   CLASXS373 &    0.33 &   10 34 29.7  +57 50 58.2 &  0.69 &                       22.70, 22.00, 21.20, 20.80, 20.30 &   9.80 &   10.07$^{+0.06}_{-0.32}$ &  0.57 \\
   CLASXS441 &    0.39 &   10 34 56.2  +57 47 24.5 &  0.85 &                       23.90, 23.00, 22.10, 21.70, 21.20 &   20.0 &   10.26$^{+0.22}_{-0.38}$ &  2.55 \\
   CLASXS448 &    0.62 &   10 34 57.9  +57 37 56.1 &  0.42 &                       25.30, 24.70, 23.90, 23.00, 22.90 &   13.0 &   10.22$^{+0.13}_{-0.17}$ &  7.49 \\
   CLASXS517 &    0.62 &   10 35 51.0  +57 43 33.0 &  0.28 &                       21.10, 21.10, 20.90, 20.50, 20.20 &   73.0 &    9.92$^{+0.04}_{-0.03}$ &  2.65 \\
   CLASXS522 &    0.51 &   10 36 04.2  +57 47 48.3 &  0.17 &                       23.60, 23.20, 22.50, 22.20, 21.70 &   1.90 &    9.74$^{+0.36}_{-0.34}$ &  0.32 \\
\enddata
\tablecomments{ Col.(1): Source name. Col.(2): Redshift. Col.(3): RA and DEC of the X-ray target. Col.(4): The distance in arcsec of
the optical counterpart from the X-ray target. Col.(5): Optical/near-IR photometry in
AB system. The photometric bands for each field are listed in  Table~\ref{xrayfields}. Col.(6): The observed-frame
2-8 KeV X-ray flux. Col.(7): The stellar mass. Col.(8): The hard X-ray to $R$-band flux ratio. }
\end{deluxetable}
\clearpage         
\end{landscape}


\begin{thebibliography}{}

\bibitem[Alexander et al.(2003)]{Alexander03} Alexander, D.~M., et al.\ 2003, \aj, 126, 539 

\bibitem[Alonso-Herrero et al.(2008)]{Alonso-Herrero08} Alonso-Herrero, A., P{\'e}rez-Gonz{\'a}lez, P.~G., Rieke, G.~H., Alexander, D.~M., Rigby, J.~R., Papovich, C., Donley, J.~L., \& Rigopoulou, D.\ 2008, \apj, 677, 127 


\bibitem[Aller \& Richstone(2002)]{Aller02} Aller, M.~C., \& Richstone, D.\ 2002, \aj, 124, 3035 

\bibitem[Barger et al.(2002)]{Barger02} Barger, A.~J., Cowie, L.~L., Brandt, W.~N., Capak, P., 
Garmire, G.~P., Hornschemeier, A.~E., Steffen, A.~T., \& Wehner, E.~H.\ 2002, \aj, 124, 1839 

\bibitem[Barger et al.(2003)]{Barger03} Barger, A.~J., et al.\ 2003, \aj, 126, 632 

\bibitem[Barger et al.(2005)]{Barger05} Barger, A.~J., Cowie, L.~L., Mushotzky, R.~F., Yang, Y., Wang, W.-H., Steffen, A.~T., \& Capak, P.\ 2005, \aj, 129, 578 

\bibitem[Barth et al.(2005)]{Barth05} Barth, A.~J., Greene, J.~E., \& Ho, L.~C.\ 2005, \apjl, 619, L151 

\bibitem[Best et al.(2005)]{Best05} Best, P.~N., Kauffmann, G., Heckman, T.~M., Brinchmann, J., Charlot, S., Ivezi{\'c}, {\v Z}., \& White, S.~D.~M.\ 2005, \mnras, 362, 25 


\bibitem[Bruzual \& Charlot(2003)]{Bruzual03} Bruzual, G., \& Charlot, S.\ 2003, \mnras, 344, 1000 


%\bibitem[Broos et al.(2002)]{Broos02} Broos, P. S., Townsley, L. K., Getman,
%K., \& Bauer, F. E.\ 2002, ACIS Extract, An ACIS Point Source Extraction
%Package (University Park: The Pennsylvania State Univ.)
%\url{http://www.astro.psu.edu/xray/docs/TARA/ae\_users\_guide.html}

\bibitem[Bundy et al.(2006)]{Bundy06} Bundy, K., et al.\ 2006, \apj, 651, 120 

\bibitem[Bundy et al.(2007)]{Bundy07} Bundy, K., et al.\ 2007, ArXiv e-prints, 710, arXiv:0710.2105 

\bibitem[Capak et al.(2004)]{Capak04} Capak, P., et al.\ 2004, \aj, 127, 180 

\bibitem[Carollo et al.(1998)]{Carollo98} Carollo, C.~M., Stiavelli, M., \& Mack, J.\ 1998, \aj, 116, 68 

\bibitem[Cassata et al.(2005)]{Cassata05} Cassata, P., et al.\ 2005, \mnras, 357, 903 

%\bibitem[Cash(1979)]{Cash79} Cash, W.\ 1979, \apj, 228, 939 

\bibitem[Chabrier(2003)]{Chabrier03} Chabrier, G.\ 2003, \pasp, 115, 763 

\bibitem[Chiappetti et al.(2005)]{Chiappetti05} Chiappetti, L., et al.\ 2005, \aap, 439, 413 

\bibitem[Coil et al.(2004)]{Coil04} Coil, A.~L., Newman, J.~A., Kaiser, N., Davis, M., Ma, C.-P., 
Kocevski, D.~D., \& Koo, D.~C.\ 2004, \apj, 617, 765 


\bibitem[Croton et al.(2006)]{Croton06} Croton, D.~J., et al.\ 2006, \mnras, 365, 11 

\bibitem[Cowie et al.(1996)]{Cowie96} Cowie, L.~L., Songaila, A., Hu, E.~M., \& Cohen, J.~G.\ 1996, \aj, 112, 839 

\bibitem[Cowie et al.(2002)]{Cowie02} Cowie, L.~L., Garmire, G.~P., Bautz, M.~W., Barger, A.~J., Brandt, W.~N., \& Hornschemeier, A.~E.\ 2002, \apjl, 566, L5 

\bibitem[Davis et al.(2003)]{Davis03} Davis, M., et al.\ 2003, \procspie, 4834, 161 

\bibitem[Davis et al.(2007)]{Davis07} Davis, M., et al.\ 2007, \apjl, 660, L1 

\bibitem[Decarli et al.(2007)]{Decarli07} Decarli, R., Gavazzi, G., Arosio, I., Cortese, L., Boselli, A., Bonfanti, C., \& Colpi, M.\ 2007, \mnras, 381, 136 


\bibitem[Dong et al.(2007)]{Dong07} Dong, X., et al.\ 2007, 
\apj, 657, 700 

%\bibitem[Di Matteo et al.(2005)]{DiMatteo05} Di Matteo, T., Springel, V., \& Hernquist, L.\ 2005, \nat, 433, 604 

%\bibitem[Fabian(2004)]{Fabian04} Fabian, A.~C.\ 2004, Coevolution of Black Holes and Galaxies, 446 

\bibitem[Favata et al.(2004)]{Favata04} Favata, M., Hughes, S.~A., \& Holz, D.~E.\ 2004, \apjl, 607, L5 

%\bibitem[Fan et al.(2004)]{Fan04} Fan, X., et al.\ 2004, \aj, 128, 515 

\bibitem[Ferrarese \& Merritt(2000)]{Ferrarese00} Ferrarese, L., \& Merritt, D.\ 2000, \apjl, 539, L9 

\bibitem[Ferrarese et al.(2006)]{Ferrarese06} Ferrarese, L., et al.\ 2006, \apjl, 644, L21 

\bibitem[Filippenko \& Sargent(1989)]{Filippenko89} Filippenko, A.~V., \& Sargent, W.~L.~W.\ 1989, \apjl, 342, L11 

\bibitem[Gallo et al.(2007)]{Gallo07} Gallo, E., Treu, T., Jacob, J., Woo, J.-H., Marshall, P., \& Antonucci, R.\ 2007, ArXiv e-prints, 711, arXiv:0711.2073 

\bibitem[Gebhardt et al.(2000)]{Gebhardt00} Gebhardt, K., et al.\ 2000, \apjl, 539, L13 
 
\bibitem[Gebhardt et al.(2001)]{Gebhardt01} Gebhardt, K., et al.\ 2001, \aj, 122, 2469 

\bibitem[Georgakakis et al.(2008)]{Georgakakis08} Georgakakis, A., et al.\ 2008, arXiv0801.2160

%\bibitem[George et al.(2000)]{George00} George, I.~M., et al.\ 2000, \apj, 531, 52 

%\bibitem[Getman et al.(2005)]{Getman05} Getman, K.~V., et al.\ 2005, \apjs, 160, 319

\bibitem[Ghosh et al.(2008)]{Ghosh08} Ghosh, H., Mathur, S., Fiore, F., \& Ferrarese, L.\ 2008, ArXiv e-prints, 801, arXiv:0801.4382 

\bibitem[Greene \& Ho(2004)]{Greene04} Greene, J.~E., \& Ho, L.~C.\ 2004, \apj, 610, 722 


\bibitem[Greene \& Ho(2007b)]{Greene07b} Greene, J.~E., \& Ho, L.~C.\ 2007, \apj, 667, 131 
%%''The Mass Function of Active Black Holes in the Local Universe''

\bibitem[Greene \& Ho(2007c)]{Greene07c} Greene, J.~E., \& Ho, L.~C.\ 2007, \apj, 670, 92 
%%``A New Sample of Low-Mass Black Holes in Active Galaxies''


%\bibitem[Haardt et al.(1997)]{Haardt97} Haardt, F., Maraschi, L., \& Ghisellini, G.\ 1997, \apj, 476, 620 

\bibitem[H{\"a}ring \& Rix(2004)]{Haring04} H{\"a}ring, N., \& Rix, H.-W.\ 2004, \apjl, 604, L89 

\bibitem[Heckman et al.(2004)]{Heckman04} Heckman, T.~M., Kauffmann, G., Brinchmann, J., Charlot, S., Tremonti, C., \& White, S.~D.~M.\ 2004, \apj, 613, 109 

\bibitem[Heckman et al.(2005)]{Heckman05} Heckman, T.~M., Ptak, A., Hornschemeier, A., \& Kauffmann, G.\ 2005, \apj, 634, 161 

\bibitem[Ho et al.(1995)]{Ho95} Ho, L.~C., Filippenko, A.~V., \& Sargent, W.~L.\ 1995, \apjs, 98, 477 

\bibitem[Ho et al.(1997)]{Ho97} Ho, L.~C., Filippenko, A.~V., \& Sargent, W.~L.~W.\ 1997, \apj, 487, 568 

\bibitem[Ho(2008)]{Ho08} Ho, L.~C.\ 2008, ArXiv e-prints, 803, arXiv:0803.2268 


\bibitem[Iovino et al.(2005)]{Iovino05} Iovino, A., et al.\ 2005, \aap, 442, 423 

%\bibitem[Kalberla et al.(2005)]{Kalberla05} Kalberla, P.~M.~W., et al. \ 2005, \aap, 440, 775 

\bibitem[Kauffmann et al.(2003a)]{Kauffmann03a} Kauffmann, G., et al.\ 2003, \mnras, 346, 1055 
%''The host galaxies of active galactic nuclei''

\bibitem[Kauffmann et al.(2003b)]{Kauffmann03b} Kauffmann, G., et al.\ 2003, \mnras, 341, 54 
%''The dependence of star formation history and internal structure on stellar mass 
%for 105 low-redshift galaxies''

\bibitem[Kauffmann et al.(2003c)]{Kauffmann03c} Kauffmann, G., et al.\ 2003, \mnras, 341, 33 
%''Stellar masses and star formation histories for 105 galaxies from the Sloan Digital Sky Survey''

\bibitem[Kormendy \& Richstone(1995)]{Kormendy95} Kormendy, J., \& Richstone, D.\ 1995, \araa, 33, 581 

%\bibitem[Lawson \& Turner(1997)]{Lawson97} Lawson, A.~J., \& Turner, M.~J.~L.\ 1997, \mnras, 288, 920 

\bibitem[Laine et al.(2003)]{Laine03} Laine, S., van der Marel, R.~P., Lauer, T.~R., Postman, M., O'Dea, C.~P., \& Owen, F.~N.\ 2003, \aj, 125, 478 

\bibitem[Laird et al.(2008)]{Laird08} Laird, E.~S., et al.\ 2008, ArXiv e-prints, 809, arXiv:0809.1349 

\bibitem[Le F{\`e}vre et al.(2004)]{LeFevre04} Le F{\`e}vre, O., et al.\ 2004, \aap, 417, 839 

\bibitem[Le F{\`e}vre et al.(2005)]{LeFevre05} Le F{\`e}vre, O., et al.\ 2005, \aap, 439, 845 

\bibitem[Lodato \& Natarajan(2006)]{Lodato06} Lodato, G., \& Natarajan, P.\ 2006, \mnras, 371, 1813 

\bibitem[Lotz et al.(2008)]{Lotz08} Lotz, J.~M., et al.\ 2008, \apj, 672, 177 


\bibitem[Madau \& Rees(2001)]{Madau01} Madau, P., \& Rees, M.~J.\ 2001, \apjl, 551, L27 

\bibitem[Maiolino \& Rieke(1995)]{Maiolino95} Maiolino, R., \& Rieke, G.~H.\ 1995, \apj, 454, 95 

\bibitem[Magorrian et al.(1998)]{Magorrian98} Magorrian, J., et al.\ 1998, \aj, 115, 2285 

\bibitem[Marconi et al.(2004)]{Marconi04} Marconi, A., Risaliti, G., Gilli, R., Hunt, L.~K., Maiolino, R., \& Salvati, M.\ 2004, \mnras, 351, 169 

\bibitem[McCracken et al.(2003)]{McCracken03} McCracken, H.~J., et al.\ 2003, \aap, 410, 17 

\bibitem[McNamara \& Nulsen(2007)]{McNamara07} McNamara, B.~R., \&  Nulsen, P.~E.~J.\ 2007, \araa, 45, 117 


\bibitem[Merritt et al.(2001)]{Merritt01} Merritt, D., Ferrarese, L., \& Joseph, C.~L.\ 2001, Science, 293, 1116 

\bibitem[Merritt et al.(2004)]{Merritt04} Merritt, D., Milosavljevi{\'c}, M., Favata, M., Hughes, S.~A., \& Holz, D.~E.\ 2004, \apjl, 607, L9 


%\bibitem[Nandra \& Pounds(1994)]{Nandra94} Nandra, K., \& Pounds, K.~A.\ 1994, \mnras, 268, 405 

\bibitem[Nandra et al.(2005)]{Nandra05} Nandra, K., et al.\ 2005, \mnras, 356, 568 

\bibitem[Nandra et al.(2007)]{Nandra07} Nandra, K., et al.\  2007, \apjl, 660, L11


\bibitem[P{\'e}rez-Gonz{\'a}lez et al.(2008)]{Perez-Gonzalez08} P{\'e}rez-Gonz{\'a}lez, P.~G., et al.\ 2008, \apj, 675, 234 


\bibitem[Radovich et al.(2004)]{Radovich04} Radovich, M., et al.\ 2004, \aap, 417, 51 

\bibitem[Ranalli et al.(2003)]{Ranalli03} Ranalli, P., Comastri, A., \& Setti, G.\ 2003, \aap, 399, 39 

\bibitem[Risaliti et al.(1999)]{Risaliti99} Risaliti, G., Maiolino, R., \& Salvati, M.\ 1999, \apj, 522, 157 

\bibitem[Rossa et al.(2006)]{Rossa06} Rossa, J., van der Marel, R.~P., B{\"o}ker, T., Gerssen, J., Ho, L.~C., Rix, H.-W., Shields, J.~C., \& Walcher, C.-J.\ 2006, \aj, 132, 1074 


%\bibitem[Saez et al.(2008)]{Saez08} Saez, C., Chartas, G., Brandt, W.~N., Lehmer, B.~D., Bauer, F.~E., Dai, X., \& Garmire, G.~P.\ 2008, ArXiv e-prints, 801, arXiv:0801.3599 

\bibitem[Satyapal et al.(2007)]{Satyapal07} Satyapal, S., Vega, 
D., Heckman, T., O'Halloran, B., \& Dudik, R.\ 2007, \apjl, 663, L9 

\bibitem[Satyapal et al.(2008)]{Satyapal08} Satyapal, S., Vega, 
D., Dudik, R.~P., Abel, N.~P., \& Heckman, T.\ 2008, \apj, 677, 926 



\bibitem[Schmidt(1968)]{Schmidt68} Schmidt, M.\ 1968, ApJ, 151, 393 

\bibitem[Shankar et al.(2004)]{Shankar04} Shankar, F., Salucci, P., Granato, G.~L., De Zotti, G., \& Danese, L.\ 2004, \mnras, 354, 1020 

\bibitem[Shi et al.(2006)]{Shi06} Shi, Y., Rieke, G.~H., Papovich, C., P{\'e}rez-Gonz{\'a}lez, P.~G., \& Le Floc'h, E.\ 2006, \apj, 645, 199 

\bibitem[Shields et al.(2008)]{Shields08} Shields, J.~C., 
Walcher, C.~J., Boeker, T., Ho, L.~C., Rix, H.-W., 
\& van der Marel, R.~P.\ 2008, ArXiv e-prints, 804, arXiv:0804.4024 

\bibitem[Sivakoff et al.(2008)]{Sivakoff08} Sivakoff, G.~R., Martini, P., Zabludoff, A.~I., Kelson, D.~D., 
\& Mulchaey, J.~S.\ 2008, ArXiv e-prints, 804, arXiv:0804.3797 

\bibitem[Steffen et al.(2004)]{Steffen04} Steffen, A.~T., Barger, A.~J., Capak, P., Cowie, L.~L., 
Mushotzky, R.~F., \& Yang, Y.\ 2004, \aj, 128, 1483 

\bibitem[Szokoly et al.(2004)]{Szokoly04} Szokoly, G.~P., et al.\ 2004, \apjs, 155, 271 




\bibitem[Valluri et al.(2005)]{Valluri05} Valluri, M., Ferrarese, L., Merritt, D., \& Joseph, C.~L.\ 2005, \apj, 628, 137 

\bibitem[Verolme et al.(2002)]{Verolme02} Verolme, E.~K., et al.\ 2002, \mnras, 335, 517 

\bibitem[Volonteri et al.(2008)]{Volonteri08} Volonteri, M., Lodato, G., \& Natarajan, P.\ 2008, \mnras, 383, 1079 

\bibitem[Wehner \& Harris(2006)]{Wehner06} Wehner, E.~H., \& Harris, W.~E.\ 2006, \apjl, 644, L17 

\bibitem[Willmer et al.(2006)]{Willmer06} Willmer, C.~N.~A., et al.\ 2006, ApJ, 647, 853 

\bibitem[Wolf et al.(2004)]{Wolf04} Wolf, C., et al.\ 2004, \aap, 421, 913 

\bibitem[Yang et al.(2004)]{Yang04} Yang, Y., Mushotzky, R.~F., Steffen, A.~T., Barger, A.~J., 
\& Cowie, L.~L.\ 2004, \aj, 128, 1501 

\bibitem[Zakamska et al.(2006)]{Zakamska06} Zakamska, N.~L., et 
al.\ 2006, \aj, 132, 1496 

\bibitem[Zezas et al.(1998)]{Zezas98} Zezas, A.~L., Georgantopoulos, I., \& Ward, M.~J.\ 1998, \mnras, 301, 915 

\end{thebibliography}
\end{document}